\documentclass[aps,prx,final,10pt,twocolumn,superscriptaddress]{revtex4-1}
\usepackage{amsmath,amssymb,bm}
\usepackage{graphicx,color}
\usepackage[squaren]{SIunits}
\usepackage[english]{babel}
\usepackage{amstext}
\usepackage{amsthm}
\usepackage{latexsym}
\usepackage{array}
\usepackage{color}
\usepackage{float}
\usepackage{microtype}

\usepackage{multirow}
\usepackage{dcolumn}
\usepackage{soul}

\definecolor{url}{RGB}{0,20,160}
\usepackage[colorlinks=true,linkcolor=blue,citecolor=blue,urlcolor=url]{hyperref}
\usepackage[usenames,dvipsnames,svgnaes,table]{xcolor}
\definecolor{col1}{rgb}{0.0, 0.30, 1.0}
\definecolor{col2}{rgb}{0.9, 0.0, 0.30}

\begin{document}
\title{{Coexisted Three-component Bosons and Two-component Weyl Bosons}
in TiS, ZrSe and HfTe}

\author{Jiangxu Li}
\affiliation{Shenyang National Laboratory for Materials Science,
Institute of Metal Research, Chinese Academy of Science, School of
Materials Science and Engineering, University of Science and
Technology of China, 110016, Shenyang, China}

\author{Qing Xie}
\affiliation{Shenyang National Laboratory for Materials Science,
Institute of Metal Research, Chinese Academy of Science, School of
Materials Science and Engineering, University of Science and
Technology of China, 110016, Shenyang, China}
\affiliation{University of Chinese Academy of Sciences, Beijing,
100049, China}

\author{Sami Ullah}
\affiliation{Shenyang National Laboratory for Materials Science,
Institute of Metal Research, Chinese Academy of Science, School of
Materials Science and Engineering, University of Science and
Technology of China, 110016, Shenyang, China}
\affiliation{University of Chinese Academy of Sciences, Beijing,
100049, China}

\author{Ronghan Li}
\affiliation{Shenyang National Laboratory for Materials Science,
Institute of Metal Research, Chinese Academy of Science, School of
Materials Science and Engineering, University of Science and
Technology of China, 110016, Shenyang, China}

\author{Hui Ma}
\affiliation{Shenyang National Laboratory for Materials Science,
Institute of Metal Research, Chinese Academy of Science, School of
Materials Science and Engineering, University of Science and
Technology of China, 110016, Shenyang, China}

\author{Dianzhong Li}
\affiliation{Shenyang National Laboratory for Materials Science,
Institute of Metal Research, Chinese Academy of Science, School of
Materials Science and Engineering, University of Science and
Technology of China, 110016, Shenyang, China}

\author{Yiyi Li}
\affiliation{Shenyang National Laboratory for Materials Science,
Institute of Metal Research, Chinese Academy of Science, School of
Materials Science and Engineering, University of Science and
Technology of China, 110016, Shenyang, China}

\author{Xing-Qiu Chen}
\email[Corresponding author:]{xingqiu.chen@imr.ac.cn}
\affiliation{Shenyang National Laboratory for Materials Science,
Institute of Metal Research, Chinese Academy of Science, School of
Materials Science and Engineering, University of Science and
Technology of China, 110016, Shenyang, China}

\date{\today}

\begin{abstract}

In analogy to various fermions of electrons in topological
semimetals, topological mechanical states with two type of bosons,
Dirac and Weyl bosons, were reported in some macroscopic systems of
kHz frequency and those with a type of doubly-Weyl phonons in atomic
vibrational framework of THz frequency of solid crystals were
recently predicted. However, to date no three-component bosons of
phonon has been reported. Here, through first-principles
calculations, we have reported that the phonon spectra of the
WC-type TiS, ZrSe, and HfTe commonly host the unique triply
degenerate nodal points (TDNPs) and single two-component Weyl points
(WPs) in THz frequency. Quasiparticle excitations near TDNPs of
phonons are three-component bosons, beyond the conventional and
known classifications of Dirac, Weyl and doubly-Weyl phonons.
{Moreover, we have found that both TiS and ZrSe have five pairs of
type-I Weyl phonons and a pair of type-II Weyl phonons, whereas HfTe
only has four pairs of type-I Weyl phonons. They carry non-zero
topological charges. On the (10$\bar{1}$0) crystal surfaces, we
observe topological protected surface arc states connecting two WPs
with opposite charges, which host modes that propagate nearly in one
direction on the surface.}
\end{abstract}

\maketitle

\section{Introductions}
Topological semimetals \cite{H.Weng2016,Y.B2016,
C.K2016,A.Bansil2016,S.Rao} are one of the fast growing families in
the frontier of material sciences and condensed matter physics due
to their unique density of states, transport properties and novel
topological surface states as well as their potential for use in
quantum computers, spintronics and novel physics. It has been
well-known that topological semimetals highlight several main types
of interesting fermions in crystal solids, such as three-dimensional
(3D) Dirac cones
\cite{Z.Wang2012,Cheng.X2014,Young2012,Liu.Z2014,Xu2015,
Z.W2013,Z.K2014,Neupane2014,Du.Y2015,J.Hul,B.J2014}, Weyl
nodes\cite{S.Murakami2007,X.Wan2011,G.Xu2011,S.Y2015,Shekhar2015,S.Y_02015,
H.Weng2015,S.M2014,B.Q2015,B.Q_02015,S.-Y2015,
L.Yang2015,Y.Zhang2016,A.A2015,Xu
SY2016,Chang2016,Yang2016,Singh2012,Ruan2016}, Dirac nodal
lines\cite{Fang.C2016,Ryu2002,Heikkila2011,Burkov.A2011,Ronghan2016,
Weng.H.M2015,Yu.R2015,Kim2015,Xie.L2015,M.G.Zeng2015,Mullen2015},
triply degenerate nodal
points\cite{B.Bradlyn2016,G.W2016,H.Weng_02016,H.Weng_12016,
Zhu2016,G.Chang2016,He.J2017,Ding.H2017,H.Yang2017,J.Yu2017}, and
even beyond\cite{B.Bradlyn2016}. In addition, their realization in
crystal solids is also important because they provide the ways to
study elementary particles, which were long-sought and predicted
ones, in high-energy physics. Importantly, in similarity to various
fermions of electrons, the exciting progresses of the bosons
(vibrational phonons) have been also predicted \cite{Lu.L2013} or
observed in the 3D momentum space of solid crystals with the
topological vibrational states, such as Dirac, Weyl and line-node
phonons in photonic crystals only with macroscopic systems of kHz
frequency
\cite{Lu.L2013,Lu.L2015,Huber.S2016,Prodan.E2009,Chen.B2014,Yang.Z2015,
Wang.P2015,Xiao.M2015,Nash.L2015,Susstrunk2015,
Mousavi2015,Fleury2016,Rocklin2016,He2016,Susstrunk2016,Lifeng2017}
and, even most recently, theoretically predicted doubly-Weyl phonons
in transition-metal monosilicides with atomic vibrations at THz
frequency \cite{Zhang.T2017}. However, to date no three-component
bosons have been reported, although three-component fermions have
been experimentally discovered in the most recent work of MoP
\cite{Ding.H2017}.

The three-component bosons would possibly occur in atomic solid
crystals because three-fold degeneracy can be protected by lattice
symmetries, such as symmorphic rotation combined with mirror
symmetries and non-symmorphic symmetries, as what was already
demonstrated to be triply degenerated points of electronic fermions
in the solid crystals
\cite{B.Bradlyn2016,G.W2016,H.Weng_02016,H.Weng_12016,Zhu2016,
G.Chang2016,He.J2017,Ding.H2017}. In addition to the importance of
seeking the new type of three-component bosons, the topological
phononic states will be extremely interesting because they could
certainly enable materials to exhibit novel heat transfer, phonon
scattering and electron-phonon interactions, as well as other
properties related with vibrational modes, such as thermodynamics.
{In the first, in similarity to topological properties of electrons,
the topological effects of phonons can induce the one-way edge
phonon states (the topologically protected boundary states). These
states will conduct phonon with little or no
scattering\cite{He2016,Mousavi2015}, highlighting possible
applications for designing phononic circuits\cite{Liuduan2016}.
Utilizing the one-way edge phonon states an ideal phonon diode
\cite{Liuduan2016} with fully 100\% efficiency becomes potential in
a multi-terminal transport system. In the second, in different from
that of electrons, as one of bosons, phonons, are not limited by the
Pauli exclusion principle. This fact demonstrates that the whole
frequency zone of phonon spectrum can be physically probed. It was
even theoretically demonstrated that the chiral phonons excited by
polarized photons can be detected by a valley phonon Hall effect in
monolayer hexagonal lattices \cite{Zhanglifa2015}. Within this
context, through first-principles calculations we report on the
novel coexistence of the triply degenerate nodal points (TDNPs) and
type-I and type-II Weyl nodes (WPs) of phonons in three compounds of
TiS, ZrSe and HfTe. Interestingly, these three materials
simultaneously still exhibit three-component fermions and
two-component Weyl fermions from their electronic structures. The
coexistence of three-component bosons, two-component Weyl bosons,
three-component fermions and two-component Weyl fermions provide
attractive candidates to study the interplays between topological
phonons and topological fermions in the same solid crystals.}

\section{Methods}

Within the framework of the density functional theory (DFT)
\cite{Hohenberg.P1964,Kohn.W1965} and the density functional
perturbation theory (DFPT) \cite{Baroni.S2001}, we have performed
the calculations on the structural optimization, the electronic band
structures, the phonon calculations and surface electronic band
structures. Both DFT and DFPT calculations have been performed by
employing the Vienna \emph{ab initio} Simulation Package (VASP)
\cite{G.Kresse1993,G.Kresse1994,G.Kresse1996}, with the projector
augmented wave (PAW) pseudopotens \cite{P.E1994,G.Kresse1999} and
the generalized gradient approximation (GGA) within the
Perdew-Burke-Ernzerhof (PBE) exchange-correlation
functional\cite{J.P1996}. The adopted PAW-PBE pseudopotentials of
all elements treat semi-core valence electrons as valence electrons.
A very accurate optimization of structural parameters have been
calculated by minimizing the interionic forces below 0.0001
eV/\AA\,. The cut-off energy for the expansion of the wave function
into the plane waves was 500 eV. The Brillouin zone integrations
were performed on the Monkhorst-Pack k-meshes
(21$\times$21$\times$23) and were sampled with a resolution of
2$\pi$ $\times$ 0.014\AA\,$^{-1}$. The band structures, either with
or without the inclusion of spin-orbit coupling (SOC), have been
performed by the Gaussian smearing method with a width of smearing
at 0.01 eV. Furthermore, the tight-binding (TB) through Green's
function methodology \cite{M.P1985,Weng2014,Weng2015} were used to
investigate the surface states. We have calculated the Hamiltonian
of tight-binding (TB) approach through maximally-localized Wannier
functions (MLWFs) \cite{N.Marzari1997,I.Souza2001} by using the
Wannier 90code \cite{A.A2008}. To calculate phonon dispersions,
force constants are generated based on finite displacement method
within the 4$\times$4$\times$4 supercells using the VASP code and
their dispersions have been further derived by Phononpy code
\cite{L.Chaput2011}. We have also computed the phonon dispersions by
including the SOC effect, which has been turned out to be no any
influence in them. {Furthermore, the force constants are used as the
tight-binding parameters to build the dynamic matrices. We determine
the topological charges of the WPs by using the Wilson-loop method
\cite{Soluyanov2011,add1}. The surface phonon DOSs are obtained by
using the iteration Green's function method \cite{M.P1985}.}

\section{Results and Discussions}

\subsection{Crystal structure and structural stabilities of the
\emph{MX} compounds}

\begin{figure}
\includegraphics[height=0.45\textwidth]{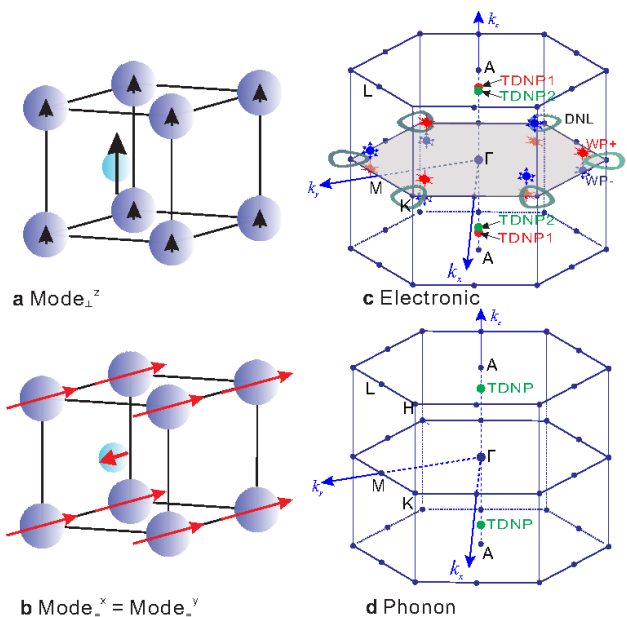}
\caption{{WC-type crystal structure and its Brilliouin zone of
\emph{MX}}(\emph{M} = Ti, Zr, Hf; \emph{X} = S, Se, Te). These
materials crystallize in the simple hexagonal crystal structure with
the space group of $P\bar{6}m2$ (No. 187). $M$ occupies the 1$a$
Wyckoff site (0, 0, 0) and $X$ locates at the 1$d$ (1/3, 2/3, 1/2)
site. Panel (\textbf{a}) shows the phonon vertical vibrational mode
(Mode$_\perp^z$) along the $k_z$ direction at the boundary -- the
high-symmetry A (0, 0, $\pi$/2) point -- of the Brilliouin zone
(BZ). Panel (\textbf{b}) denotes the phonon planar vibrational mode
(Mode$_{=}^{x}$) along the $k_x$ direction, which is two-fold
degenerate (Mode$_{=}^{x,y}$ = Mode$_{=}^{x}$ = Mode$_{=}^{y}$)
because of its C$_{3v}$ rotational symmetry. Panel (\textbf{c}) The
BZ in which the closed loops around each $K$ point denotes the Dirac
nodal lines (DNLs) of electrons around the Fermi level when SOC is
ignored. With the SOC inclusion each DNL is broken into two Weyl
points with the opposite chirality, marked as blue (WP-) and red
(WP+) balls and they coexist with the triply degenerate nodal point
(TDNP) of electronic structure (namely, three-component fermion).
Panel (\textbf{d}) shows the triply degenerate nodal point (TDNP) of
phonon dispersions ( three-component boson) along the $\Gamma$-A
direction in the BZ.} \label{fig1}
\end{figure}

Recently, the type of WC-type materials (Fig. \ref{fig1}(a)),
including ZrTe, TaN, MoP and WC, has been theoretically reported to
host the coexistence of the TDNPs and WPs in their electronic
structures. This type of coexisted fermions of electronic TDNPs and
WPs have been recently confirmed in MoP \cite{Ding.H2017}. We
further extended this family by proposing eight compounds (TiS,
TiSe, TiTe, ZrS, ZrSe, HfS, HfSe and HfTe), which are isoelectronic
and isostructural to ZrTe. Among these compounds, five compounds of
TiS, ZrS, ZrSe$_{0.90}$, and Hf$_{0.92}$Se as well as ZrTe were
experimentally reported to have the same WC-type structure
\cite{Hahn_01959,Harry1957,Steiger1970,Hahn1959,Schewe1994,
Sodeck1979,G.O2001,G.O2014}. {No any experimental data is available
for the remaining four compounds of TiSe, TiTe, HfS, and HfTe. Here,
in order to systematically investigate their electronic structures
and phonon spectra and to compare their differences, we have
considered that all these nine compounds crystallize in the same
WC-type structure.} For five experimentally known compounds TiS,
ZrS, ZrSe, ZrTe and HfSe, our DFT calculations yield the good
agreement of their equilibrium lattice parameters with the
experimental data (see supplementary Table S1). Their enthalpies of
formation are derived in supplementary Table S1, indicating their
stabilities in the thermodynamics and their phonon dispersions have
no any imaginary frequencies, revealing the stabilities in the
atomic mechanical vibrations.

\begin{figure*}[!htp]
\vspace{0.5cm}
\includegraphics[height=0.37\textwidth]{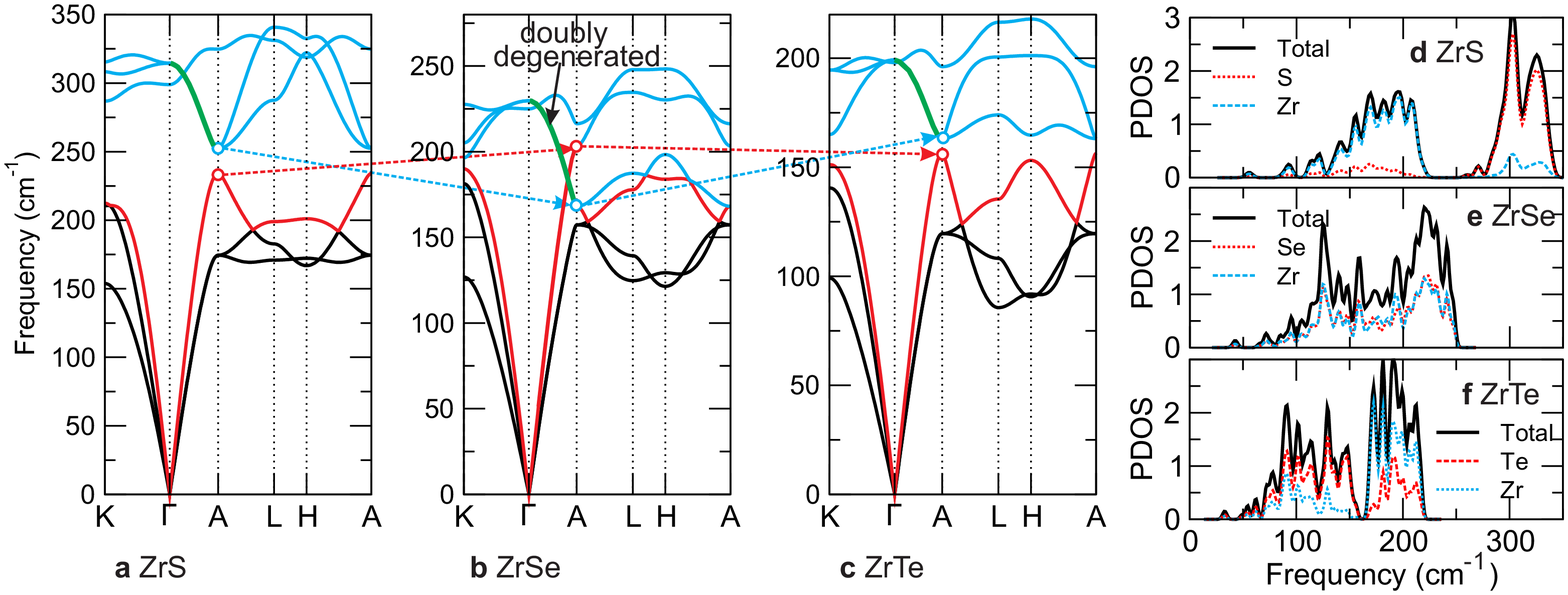}
\caption{{Phonon spectra of ZrS, ZrSe and ZrTe.} Panels (\textbf{a},
\textbf{b}, and \textbf{c}): DFT-derived phonon dispersions of ZrS,
ZrSe and ZrTe, respectively. Panels (\textbf{d}, \textbf{e}, and
\textbf{f}): DFT-derived total and partial phonon densities of
states (PDOS) of ZrS, ZrSe, and ZrTe, respectively.} \label{fig2}
\end{figure*}

\subsection{Three-component fermions and two-component Weyl
fermions in the electronic structures}

{We have elucidated the electronic band structures of these nine
compounds. Interestingly, they are in similarity to the case of ZrTe
in Ref. \onlinecite{H.Weng_12016}. As an example, the electronic
band structure of ZrSe is given in the supplementary Fig. S1,
indicating the coexisted fermions, TDNPs and WPs, whose coordinators
are further compiled in Fig. \ref{fig1}c. Of course, the similar
electronic behaviors can be observed for other compounds. But, TiS
is unique. Because of its rather weak spin-orbit coupling (SOC)
effect, TiS exhibits the coexistence of the six DNLs and the two
six-fold degenerate nodal points of its electronic structure in the
BZ. This situation is exactly what happens for other eight compounds
when the SOC effect is ignored. Basically, the appearance of these
two types of fermions, TDNPs and WPs, in this family share the same
physics, as previously discussed for ZrTe\cite{H.Weng_12016}. The
details of their electronic structures and their topologically
protected non-trivial surface states refer to the supplementary
Figs. S1, S2, S3, and S4 as well as the corresponding supplementary
texts.}

\subsection{Triply degenerate nodal points (TDNPs) of the
phonons in TiS, ZrSe and HfTe}

We have found that the presence of the triply degenerate nodal
points (TDNPs) of the phonons in three compounds of TiS, ZrSe and
HfTe after a systemical analysis of their phonon dispersions
(supplementary Fig. S5). Because each primitive cell contains two
atoms (Fig. \ref{fig1}(a)), their phonon dispersions have six
branches consisting of three acoustic and three optical ones,
respectively. As compared with the computed phonon dispersions in
Fig. \ref{fig2}(a, b, and c) and their phonon densities of states in
Fig. \ref{fig2}(d, e, and f) of the isoelectronic ZrS, ZrSe and ZrTe
compounds, a well-separated acoustic-optical gap can be observed in
both ZrS and ZrTe with the smallest direct gap at the A point (0, 0,
$\pi$/2) on the boundary of the BZ. The specified analysis uncovered
that for both ZrS and ZrTe compounds the top phonon band of the gap
at the A point is comprised with the doubly degenerate vibrational
mode of phonons in which both Zr and S (or Te) atoms, oppositely and
collinearly, displace along either $x$ or $y$ direction
(Mode$_{=}^{x,y}$ as marked in Fig. \ref{fig1}(b)). The vibrational
amplitude of the Mode$_{=}^{x,y}$ are contributed nearly 100\% by
the Zr atom, rather than by S (or Te) atoms. The bottom phononic
band of the gap at the A point is a singlet state originated from
the vibrational mode at which both Zr and S (or Te) atoms
collinearly move in the same $k_z$ direction (Mode$_{\perp}^{z}$ as
marked in Fig. \ref{fig1}(a)). But its amplitude of this
Mode$_{\perp}^{z}$ are almost fully dominated by the displacement of
S (or Te) atoms.

\begin{figure*}[!htp]
\includegraphics[height=0.37\textwidth]{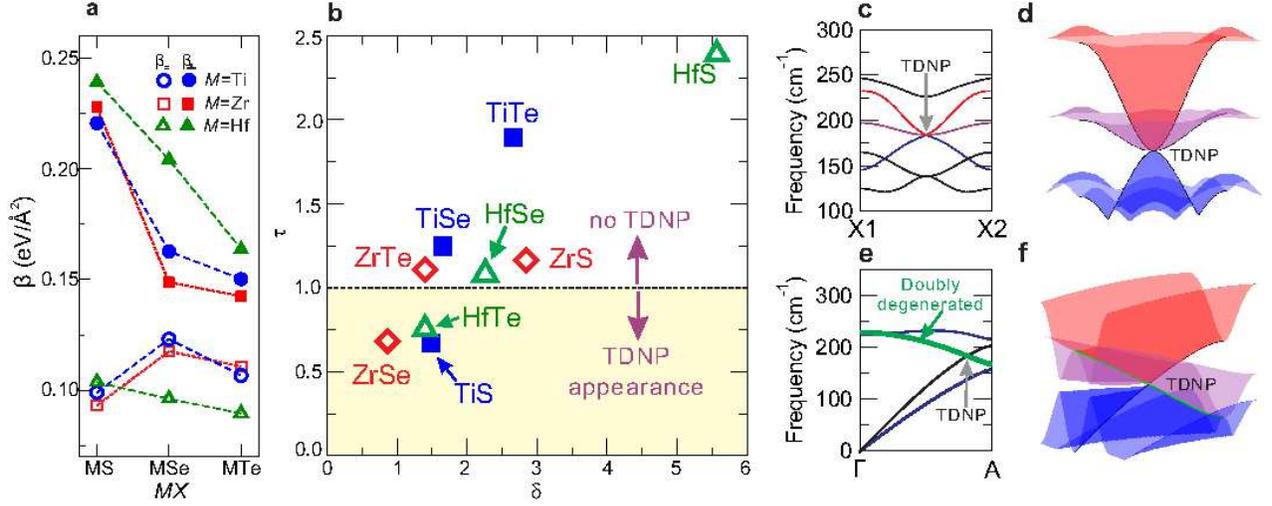}
\caption{{{Second-order force constant at the A point and the
dimensionless ratio $\tau$ of \emph{MX}}. Panel (\textbf{a}):
DFT-derived second-order force constant at the A point for both the
two-fold degenerate planar vibrational Mode$_=^{x,y}$ and the
vibrational Mode$_{\perp}^z$. Panel (\textbf{b}): The derived
parameter $\tau$ from Equ. (1) as a function of the $\delta$ value,
as defined in the main text, for all nine compounds. Panels
(\textbf{c} and \textbf{e}): DFT-derived phonon dispersions to
elucidate phonon TDNPs of ZrSe along the X1 (-$\pi$/2, 0, 0) to X2
($\pi$/2, 0, 0) and $\Gamma$-A directions, respectively. Panels
(\textbf{d} and \textbf{f}): Zoom-in 3D visualization of phonon
TDNPs in the $k_z$ = 0 and $k_y$ = 0 planes, respectively.}}
\label{fig3}
\end{figure*}

In contrast to both ZrS and ZrTe in Fig. \ref{fig2}, the case of
ZrSe shows no acoustic-optical gap (Fig. \ref{fig2}(b)), as
illustrated by its phonon density of states in \ref{fig2}(e)). It
has been noted that the planar Mode$_{=}^{x,y}$ at the A point
becomes now lower in frequency than the Mode$_{\perp}^{z}$.
Accordingly, this fact corresponds to the occurrence of phonon band
inversion at the A point. It means the unusual fact that around A
point the optical phonon bands inverts below the acoustic band which
normally should have a lower frequency. Physically, within the
(quasi)harmonic approximation the vibrational frequency, $\omega$,
have to be proportional to $\sqrt{\beta/m}$ at the boundary of the
BZ. Here, $\beta$ is the second-order force constant -- the second
derivative of the energy following a given vibrational mode as a
function of the displacement and $m$ the atomic mass. Therefore, as
seen in Fig. \ref{fig2}(b) for ZrSe the occurrence of the phonon
band inversion at the boundary A point is certainly induced by both
$\beta$ and $m$ which are determined by the planar Mode$_{=}^{x,y}$
and the Mode$_{\perp}^{z}$ at the A point. Following this
consideration, we have defined the dimensionless ratio $\tau$ as
follows,
\begin{equation}
\tau = \frac{\sqrt{\beta_=/m_=}}{\sqrt{\beta_\perp/m_\perp}},
\end{equation}
where $\tau$ specifies the comparison between the frequencies of
both Mode$_{=}^{x,y}$ and Mode$_{\perp}^z$. With $\tau$ $>$ 1 the
material shows no band inversion, thereby indicating no TDNPs. When
$\tau$ $<$ 1 implies the appearance of the phonon band inversion
with the TDNPs in the acoustic and optical gap. With such a
definition, we further plot the $\beta$ with the sequence of ZrS,
ZrSe and ZrTe in Fig. \ref{fig3}(a). It has been found that, only
with the second-order force constants of $\beta_{=}$ and
$\beta_{\perp}$ (Fig. \ref{fig3}(a)) it is not enough to induce the
phonon band inversion. This fact is in agreement with the Eq. (1)
although the $\beta_{=}$-$\beta_{\perp}$ difference is the smallest
in ZrSe among them in Fig. \ref{fig3}(a). Furthermore, for all nine
compounds in this family we compiled their $\tau$ values as a
function of the ratio ($\delta$) of the atomic masses related with
Mode$_{=}^{x,y}$ over Mode$_{\perp}^z$ (namely, $\delta$ =
$m$(Mode$_{=}^{x,y}$)/$m$(Mode$_{\perp}^{z}$)) in Fig.
\ref{fig3}(b). {With increasing the ratio of the atomic masses, the
$\tau$ value increases in a nearly linear manner. This implies that,
if the atomic masses of constituents in a targeted material highly
differ, the possibility to have TDNPs in the acoustic and optical
gap of its phonon dispersion is extremely low. However, if they have
the comparable atomic masses with the $\delta$ ratio close to 1 the
possibility to have TDNPs is high in the acoustic and optical gap.
Following this model, we have further uncovered that, because the
$\tau$ value is smaller than 1, both TiS and HfTe have similar
property as what ZrSe does (Fig. \ref{fig3}(b)). The findings for
both TiS and HfTe are in accordance with the DFT-derived phonon
dispersions in supplementary Fig. S5. However, there is no TDNP in
the acoustic and optical gap of the other members. These facts imply
that in these materials the difference between the atomic masses of
constituents in compound plays a key role in inducing the phonon
band inversion for the appearance of TDNPs in the acoustic and
optical gap, as seen for three cases of TiS, ZrSe and HfTe whose
$\delta$ value are all around 1.}

Importantly, as accompanying with the occurrence of the phonon band
inversion, the TDNPs, featured by a linear crossing of the
frequencies between the acoustic and optical bands, unavoidably
appear at (0, 0, $k_z$ = $\pm$0.40769) along the $\Gamma$-A
direction in the BZ (Fig. \ref{fig2}(b) and Fig. \ref{fig3}) for
ZrSe. Their appearance of the TDNPs in the acoustic and optical gap
is indeed protected by the C$_{3z}$ rotation and mirror symmetries
along the $\Gamma$-A direction because C$_{3z}$ allows the
coexistence of two-fold (Mode$_{=}^{x,y}$) and one-fold
(Mode$_{\perp}^Z$) representations, in similarity to their
electronic band structures as discussed above. To elucidate the
underlying mechanism of the phonon TDNPs in the acoustic and optical
gap, it still needs to be emphasized that, on the one hand, the
rotation and mirror symmetries substantially provide the
prerequisite to produce these two competing modes (two-fold
Mode$_{=}^{x,y}$ and one-fold Mode$_{\perp}^Z$) and, on the other
hand, the comparable atomic masses of constituent elements are
another ingredient to trigger the phononic band inversion. Of
course, at this TDNP it still implies that the planar
Mode$_{=}^{x,y}$ and the Mode$_{\perp}^Z$ at (0, 0, $k_z$ =
$\pm$0.40769) locate at the strictly same frequency of 183.9
cm$^{-1}$. The TDNPs locate at (0, 0, $k_z$ = $\pm$0.40382) with the
frequency of 293.4 cm $^{-1}$ for TiS and at (0, 0, $k_z$ =
$\pm$0.43045) with the frequency of 133.3 cm $^{-1}$ for HfTe. To
elucidate the 3D TDNP shape of ZrSe, we also plot the zoom-in
dispersions on both $k_{z}$ = 0 and $k_y$ = 0 planes of BZ in Fig.
\ref{fig3}. From both Fig. \ref{fig3}(c) and \ref{fig3}(d) in the
$k_{z}$ = 0 the TDNP in the acoustic and optical gap can be clearly
visualized to have an isotropic shape. However, in the $k_y$ = 0
plane the phonon bands around the TDNPs are highly complex with the
helicoid shape (Fig. \ref{fig3}(e) and \ref{fig3}(f)).

\subsection{Two-component Weyl phonons in TiS, ZrSe and HfTe}

\begin{figure*}[!htp]
\includegraphics[height=0.7\textwidth]{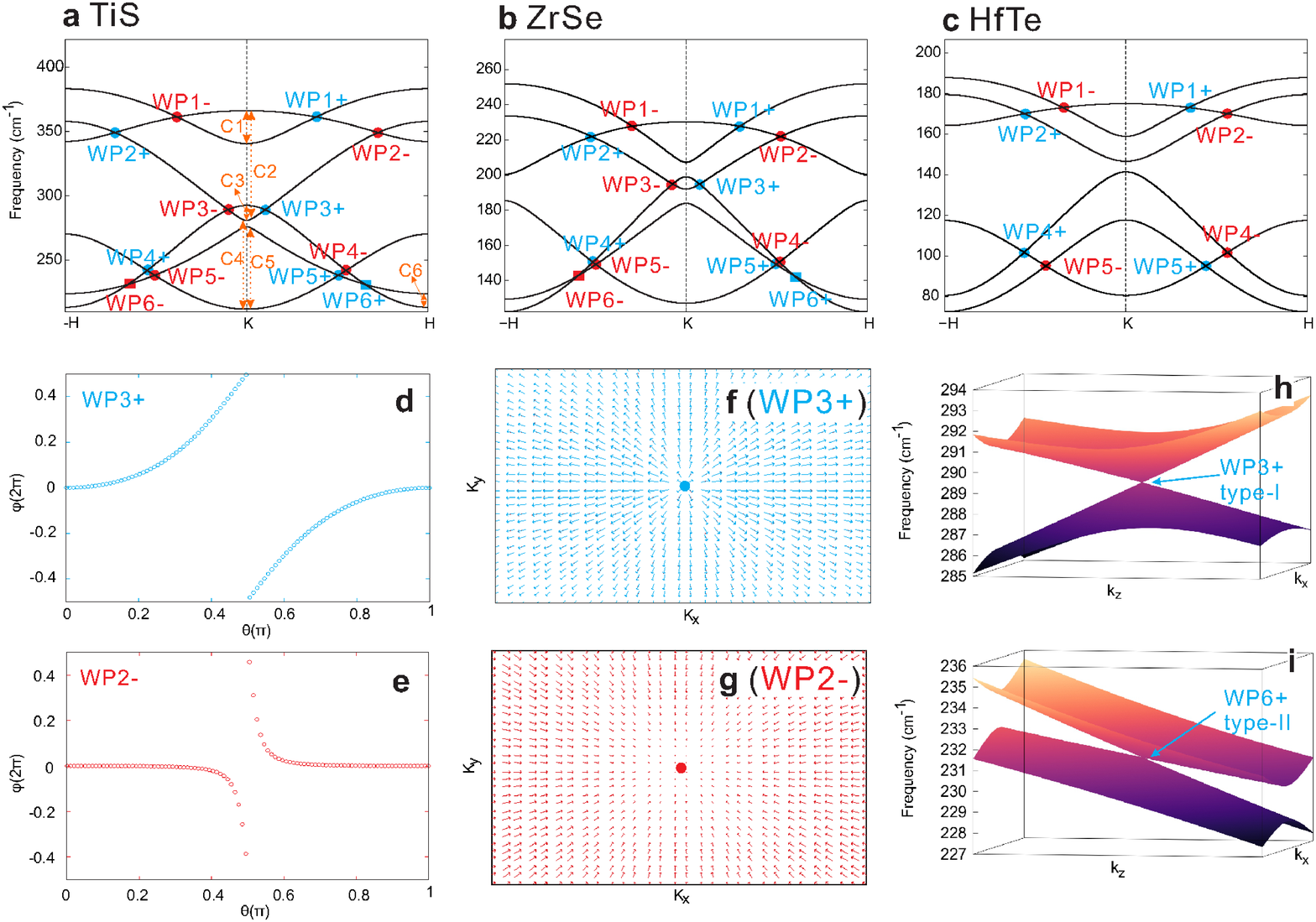}
\caption{{{Topological phonons of TiS, ZrSe and HfTe}. Panels
(\textbf{a}, \textbf{b} and \textbf{c}): DFT-derived phonon
dispersions along K to H for TiS, ZrSe and HfTe, respectively.
Panels (\textbf{d} and \textbf{e}) show the Wannnier center
evolutions around positive charge WP3+ and negative charge WP2-
nodes for TiS, respectively. Panels (\textbf{g} and \textbf{f})
denotes the Berry curvature distributions around WP3+ and WP2- Weyl
nodes for TiS. Panels (\textbf{h} and \textbf{i}) the phonon
dispersions around a type-I WP3+ and a type-II WP6+ weyl node for
TiS, respectively. Noted that the symbols of WP1$\sim$WP6 are the
Weyl nodes, the symbols of C1 to C6 refer to six different band
crossings, and the signs of + and - denote the topological positive
and negative charges, respectively.}} \label{fig4}
\end{figure*}

{Besides the existence of the TDNPs in TiS, ZrSe and HfTe, the
calculations revealed the occurrence of the two-component Weyl nodes
(WPs) in their phonon spectra. As evidenced in Fig. \ref{fig4}(a)
for TiS, the phonon bands have five different band crossings (from
C1 to C5) at the high-symmetry K point and a band crossing at the H
point. In particular, because these crossings are not constrained by
any mirror symmetry, they result in the appearance of six pairs of
WPs (Table \ref{tab1}). Among them, the band crossings from C1 to C5
confirm the five pairs of type-I WPs from WP1 to WP5 and the C6
crossing gives rise to the sixth pair of type-II WP6 one. The phonon
dispersions of type-I and type-II WPs are shown in Fig. \ref{fig4}h
and Fig. \ref{fig4}i, respectively. To identify their topological
non-trivial properties, we have calculated the topological charge of
each Weyl node, which is defined by the integration of Berry
curvature using a closed surface surrounding a node within the
framework of the Wilson-loop method \cite{Soluyanov2011,add1}. For
instance, Fig. \ref{fig4}(d and e) shows the Wannier center
evolutions around WP3+ and WP2- with the topological positive and
negative charges, respectively. Their corresponding Berry curvatures
are shown in Fig. \ref{fig4}(f and g), indicating that the positive
and negative charges, WP3+ and WP2-, have different winding
directions of their Berry curvatures. Furthermore, we determine the
charges of all the WPs of TiS in Table I. In similarity, ZrSe shares
the same six pairs of WPs (5 pairs for type-I ones and a pair for
type-II one) in Fig. \ref{fig4}b whereas HfTe only has four pairs of
type-I WPs in Fig. \ref{fig4}c, whose coordinators are given in
Table \ref{tab1}. This difference is mainly because in HfTe the
phonon dispersions from K to H are lacking of two band crossings, C3
at K and C6 at H.}

\begin{figure*}
\includegraphics[width=1.0\textwidth]{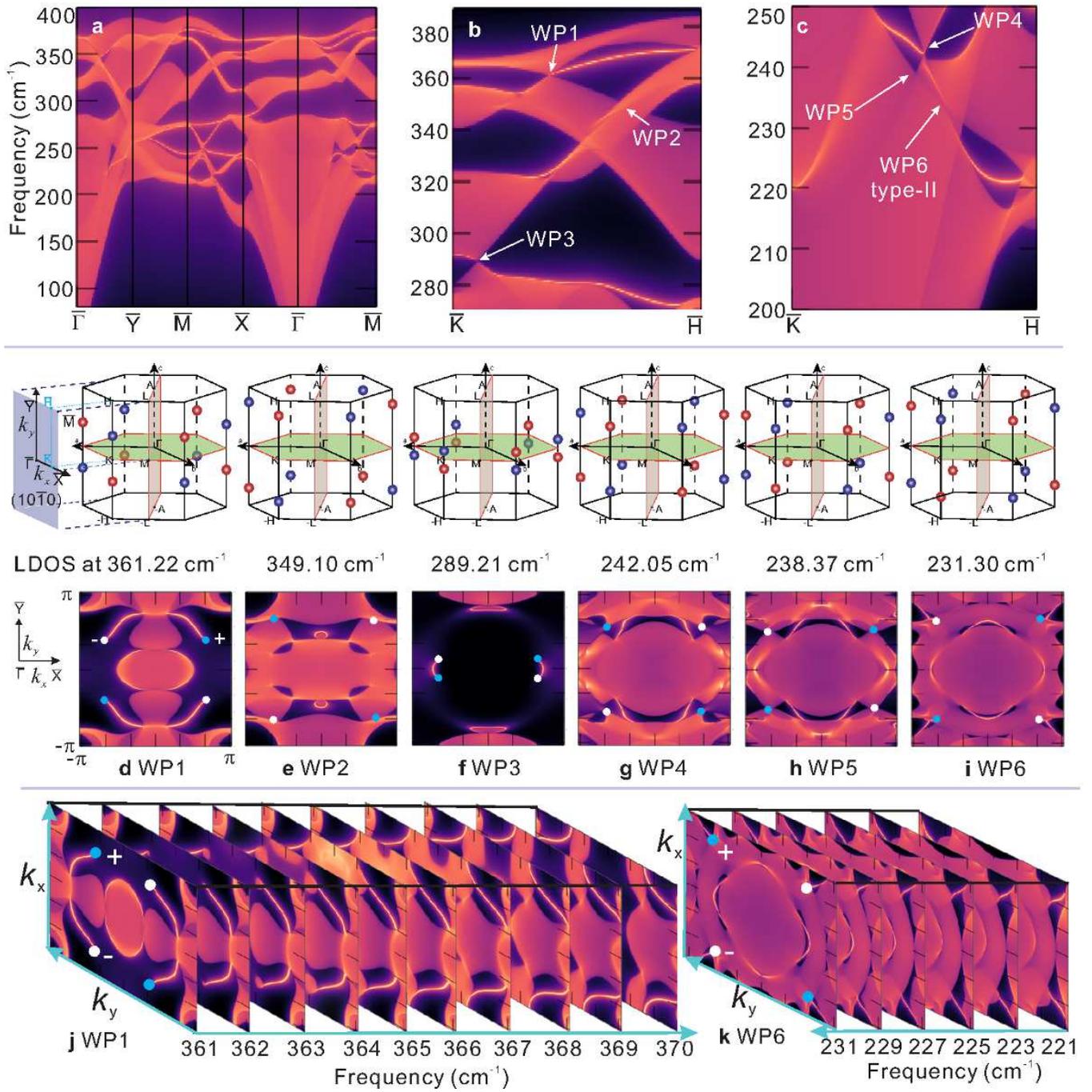}
\caption{{{The surface phonon spectra and the surface phonon
densities of states (PDOSs) of the (10$\bar{1}$0) surface of TiS}.
Panels (a, b, c): the surface phonon spectra along the high-symmetry
lines in panel (a) and along the defined $\bar{K}$-$\bar{H}$ line of
the BZ in the ((10$\bar{1}$0) surface. Panel (d to e): the surface
PDOSs at the six frequencies that the six pairs of Weyl nodes have
and the projections of these bulk WPs are marked as solid blue
(positive topological charge) and white (negative topological
charge) circs in each panel. The surface opening arc states connect
two WPs with opposite charges can be visualized in panels (d to i).
Panels j and k: the frequency-dependent evolutions of the arc states
connecting the type-I WP1 and the type-II WP6 on the the
(10$\bar{1}$0) surface of TiS, respectively.}} \label{fig5}
\end{figure*}

\begin{figure*}
\includegraphics[width=1.0\textwidth]{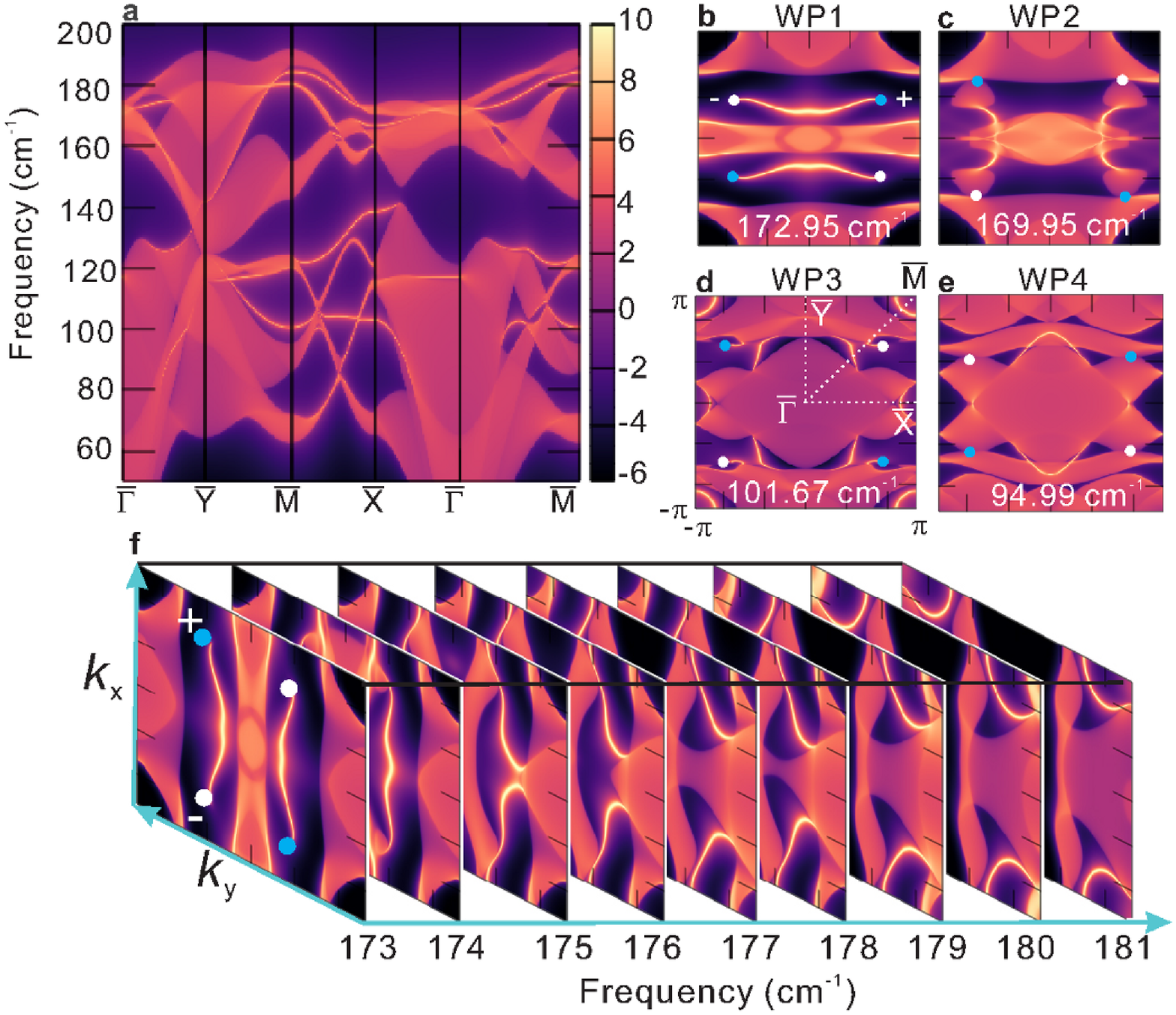}
\caption{{{The surface phonon dispersion and its evolution of the
PDOSs on the the (10$\bar{1}$0) surface of HfTe}. Panel (a): The
surface phonon dispersion on the (10$\bar{1}$0) surface of HfTe.
Panels (b to e): The surface phonon densities of states (PDOS) at
the frequencies of four pairs of Weyl nodes (WP1, WP2, WP4 and WP5
in HfTe). The projections (solid white circs - negative charge and
solid blue circs - positive charge) of the bulk WPs (and their
symmetric counterparts) on the (10$\bar{1}$0) surface are indicated
on each figure. The surface arcs connect two WPs with opposite
charges. Panel f: the frequency-dependent evolutions of the arc
states connecting the type-I WP1 on the (10$\bar{1}$0) surface of
HfTe.}} \label{fig6}
\end{figure*}

{Certainly, the existence of these WPs gives rise to the
topologically protected non-trivial surface states (TPSSs) of the
surface phonon dispersions. As shown in Fig. \ref{fig5}(a, b and c),
we have calculated the surface phonon spectrum of the (10$\bar{1}$0)
surface of TiS along the high-symmetry momentum paths in the surface
BZ. In particular, in order to see the projections of all WPs on the
(10$\bar{1}$0) surface, we have plot the surface phonon dispersions
(Fig. \ref{fig5}(b and c)) along the $\bar{K}$-$\bar{H}$ direction,
as defined in the (10$\bar{1}$0) surface BZ (Fig. \ref{fig5}d). This
$\bar{K}$-$\bar{H}$ direction indeed is the projection of the K-H
direction in the bulk BZ. As evidenced in Fig. \ref{fig5}b, the
three type-I WPs, WP1, WP2, and WP3, are clearly demonstrated and
the other two type-I WP4 and WP5 as well as another type-II WP6 can
be apparently seen in the Fig. \ref{fig5}c. Accordingly, we have
observed the interesting TPSSs, which are typically connecting each
WP in \ref{fig5}(b and c). We further plot their 2D visualization of
their phonon density of states (PDOSs) in Fig. \ref{fig5}(d to i)
using the exact frequencies with 361.22 cm$^{-1}$ of WP1, 349.10
cm$^{-1}$ of WP2, 289.21 cm$^{-1}$ of WP3, 242.05 cm$^{-1}$ of WP4,
238.37 cm$^{-1}$ of WP5, and 231.30 cm$^{-1}$ of WP6, respectively.
Interestingly, at each frequency for the (10$\bar{1}$0) surface in
Fig. \ref{fig5}(d to i), the TPSSs featured by the broken surface
arcs connecting two WPs with opposite charges for WP1, WP2, WP3 and
WP6 can be clearly visualized. However, it is a bit difficult to
observe the broken arcs states connecting WP4 and WP5 on the
(10$\bar{1}$0) surface because they are heavily overlapped with the
projections of bulk phonon states. The case of ZrSe also exhibits
the quite similar arc states of surface phonon on its (10$\bar{1}$0)
surface (not shown here).}

{As compared with both cases of both TiS and ZrSe, HfTe exhibits
some differences. HfTe only has four pairs of type-I WPs as marked
in Fig. \ref{fig4}c and no type-II WPs. Fig. \ref{fig6} shows its
phonon spectrum of the (10$\bar{1}$0) surface and the 2D
visualizations of the PDOSs with the frequencies of 172.95 cm$^{-1}$
of WP1, 169.95 cm$^{-1}$ of WP2, 101.67 cm$^{-1}$ of WP4, and 94.99
cm$^{-1}$ of WP5, respectively. The bulk WPs are also projected onto
the (10$\bar{1}$0) surface. As shown Fig. \ref{fig6}b, the broken
arc states of the TPSSs are clearly linked to the pair of WP1 with
opposite topological charges, and only partial for both WP2 and WP4
in Fig. \ref{fig6}(c and d), and not observable for WP5 due to its
overlapping with the projected states of bulk phonon dispersions in
Fig. \ref{fig6}e. In addition, it still needs to be emphasized that
the arc states can be certainly observed on some other planes which
are paralleling to the bulk H-K direction, such as the
(01$\bar{1}$0) plane. However, note that the arc states connecting
Weyl nodes cannot be observable on the (0001) surface because, on
it, the projections of the K-H direction coincide at the same
surface momentum, and their topological charges cancel to each
other.}

\begin{table*}[!t]
\begin{center}
\caption{Weyl points at $\mathbf{k} = (\frac{1}{3}, \frac{1}{3},
k_z)$ and their frequencies $\omega$, topological charges (+ or -)
and types (type-I or type-II) of TiS, ZrSe and HfTe. \label{tab1}}
\begin{ruledtabular}
\begin{tabular}{c|cccc|cccc|cccc}
WPs & {$k_z$} & {$\omega$ (cm$^{-1}$)} & {Charge} & {Type} & {$k_z$} &  {$\omega$ (cm$^{-1}$)} & {Charge} & {Type} & {$k_z$} & {$\omega$ (cm$^{-1}$)} & {Charge} & {Type} \\
\hline
WP1& 0.1919   &   361.22  &   $+$   &  I  & 0.1485   &   227.76  &   $+$   &  I  & 0.1739   &   172.95    &   $+$   &  I  \\
WP2& 0.3628   &   349.07  &   $-$   &  I  & 0.2600   &   221.91  &   $-$   &  I  & 0.2798   &   169.95    &   $-$   &  I     \\
WP3& 0.0517   &   289.21  &   $+$   &  I  & 0.0371   &   194.93  &   $+$   &  I      \\
WP4& 0.2741   &   242.05  &   $-$   &  I  & 0.2569   &   150.52  &   $-$   &  I  & 0.2803   &   101.67    &   $-$   &  I         \\
WP5& 0.2533   &   238.37  &   $+$   &  I  & 0.2486   &   149.22  &   $+$   &  I  & 0.2205   &   94.99     &   $+$   &  I        \\
WP6& 0.3265   &   231.30  &   $+$   &  II & 0.2977   &   142.84  &   $+$   &  II      \\
\end{tabular}
\end{ruledtabular}
\end{center}
\end{table*}

\section{Discussions}

{Through the DFT-derived results, these three materials of TiS, ZrSe
and HfTe are highly attractive because of the occurrence of the
coexisted TDNPs and WPs. In the first, the TDNPs of their phonons
are interesting because (i) they provide a good platform to study
the behaviors of the basic triple degenerate boson, one of
elementary particles, in the real materials, (ii) they are highly
robust, which are locked by the threefold rotational symmetry of the
hexagonal lattices, and (iii) they exactly occur in the
optical-acoustic gap and do not overlap with other phonon bands.
Perhaps, the thermal-excited signals related with these TDNPs will
not be interfered by other vibrational modes, thereby highlight the
viable cases to experimentally probe the TDNP-related properties.}

{In the second, it is well-known that in the electronic structures
the WPs and their associated topological invariants enable the
corresponding materials to exhibit a variety of novel properties,
such as robust surface states and chiral anomaly
\cite{S.Murakami2007, X.Wan2011, G.Xu2011, S.Y2015, Shekhar2015,
S.Y_02015, H.Weng2015, S.M2014, B.Q2015, B.Q_02015, S.-Y2015,
L.Yang2015, Y.Zhang2016, A.A2015, Xu SY2016, Chang2016, Yang2016,
Singh2012, Ruan2016}. In our current cases, the existence of the
bulk phononic WPs and their robust TPSSs render them to be very
charming for possible applications, because these states can not be
backscattered. In particular, as evidenced in Fig. \ref{fig5}j the
surface broken arc states connecting a pair of WP1 nodes in TiS
exhibit an nearly one-way propagation. Its evolution further extends
and shift to the zone boundary with increasing the frequencies in a
relatively wide region of frequency in Fig. \ref{fig5}j. In
similarity, the nearly one-way arc states connecting a pair of WP1
nodes in HfTe can be clearly visualized in Fig. \ref{fig6}f.
However, the evolution of the surface arc states connecting a pair
of type-II WP6 in TiS cannot be fully visualized because most of
them are overlapped with the projections of the bulk phonon states
in Fig. \ref{fig5}k.}

\section{SUMMARY}

{Summarizing, through first-principles calculations we have revealed
that three WC-type materials of TiS, ZrSe and HfTe not only host
three-component bosons featured by TDNPs and two-component Weyl
bosons featured by WPs in their phonon spectra. In both TiS and
ZrSe, there exist six pairs of bulk WPs (five type-I nodes and one
type-II node) locating at the K-H line in the BZ, whereas in HfTe
only four pairs of type-I WPs exist. We have demonstrated that their
phonon spectra of these three cases are topological in nature,
exhibiting that the topologically protected non-trivial surface arc
states of phonons. These non-trivial states are directly linked with
various WPs with opposite chirality. Interestingly, these three
cases still exhibit three-component fermions featured by TDNPs and
six pairs of two-component Weyl fermions (WPs) in their electronic
structures of the bulk crystals. The novel coexistence of the main
features of (\emph{i}) three-component bosons, (\emph{ii})
two-component Weyl bosons, and three-component fermions, and
(\emph{iii}) two-component Weyl fermions and, in particular, both
three-component bosons and three-component fermions at the nearly
same momentum (Fig. \ref{fig1}(c and d)) along the $\Gamma$-A
direction could couple to each other through electron-phonon
interactions. They hence highlight a wonderful platform to study the
interplays between different types of topological electron
excitations and topological phonons within the atomistic scale for
potential multi-functionality quantum-mechanical properties.}

\bigskip

\vspace{0.5cm} \noindent {\bf Acknowledgments}

\noindent We thank H. M. Weng for valuable discussions. Work was
supported by the National Science Fund for Distinguished Young
Scholars (No. 51725103), by the National Natural Science Foundation
of China (Grant Nos. 51671193 and 51474202), and by the Science
Challenging Project No. TZ2016004. All calculations have been
performed on the high-performance computational cluster in the
Shenyang National University Science and Technology Park and the
National Supercomputing Center in Guangzhou (TH-2 system) with
special program for applied research of the NSFC-Guangdong Joint
Fund (the second phase) under Grant No.U1501501.

\pagebreak \widetext \clearpage
\begin{center}
\textbf{\large Supplemental Materials: ''Coexisted Three-component
and Two-component Weyl bosons in the topological semimetals of TiS,
ZrSe and HfTe''}

\vspace{1cm}

Jiangxu Li$^{1}$, Qing Xie$^{1,2}$, Sami
Ullah$^{1,2}$, Ronghan Li$^1$, Hui Ma$^1$, Dianzhong
Li$^1$, Yiyi Li$^1$, Xing-Qiu Chen$^1,\textcolor{blue}{*}$ \\

\vspace{1cm} $^1$ \emph{Shenyang National Laboratory for Materials
Science, Institute of Metal Research, Chinese Academy of Science,
School of Materials Science and Engineering, University of Science
and Technology of
China, 110016, Shenyang, China} \\
$^2$ \emph{University of Chinese Academy of Sciences, Beijing, 100049,
China}\\
\end{center}

\vspace{1cm}
\textcolor{blue}{*} Corresponding
author:{\textcolor{blue}{xingqiu.chen@imr.ac.cn} (X.-Q. C.)}


\setcounter{equation}{0}
\setcounter{figure}{0}
\setcounter{table}{0}
\setcounter{page}{1}
\makeatletter
\renewcommand{\thepage}{S\arabic{page}}
\renewcommand{\thetable}{S\arabic{table}}
\renewcommand{\theequation}{S\arabic{equation}}
\renewcommand{\thefigure}{S\arabic{figure}}
\renewcommand{\bibnumfmt}[1]{[S#1]}
\renewcommand{\citenumfont}[1]{#1}

\vspace{0.3cm}

\vspace{0.5cm}
\begin{center}
 {\bf \Large Supplementary Materials}
\end{center}
\begin{enumerate}
  \item \textbf{Table S1}: Optimized lattice parameters of $MX$
  \item \textbf{Figure S1}: Electronic structures of ZrSe
  \item \textbf{Figure S2}: Evolution of the electronic structure around the Weyl points
  (WPs) in ZrSe
  \item \textbf{Figure S3}: Surface electronic band structures
  of (0001) and (10$\bar{1}$0) surfaces of ZrSe
  \item \textbf{Figure S4}: Fermi surfaces of the (0001) and (10$\bar{1}$0) surfaces of
  ZrSe
  \item \textbf{Figure S5}: DFT-derived phonon dispersions of
  the nine $MX$ compounds
\end{enumerate}

\clearpage

\subsection{Supplementary Table S1}

We have optimized the lattice structures of nine $MX$ compounds with
the WC-type structure. Table S1 summarizes all optimized lattice
constants as compared with the available experimental data. Among
these nine compounds, five compounds of TiS, ZrS, ZrSe$_{0.90}$, and
Hf$_{0.92}$Se as well as ZrTe were experimentally reported to have
the same WC-type structure
\cite{Hahn_01959,Harry1957,Steiger1970,Hahn1959,Schewe1994,Sodeck1979,G.O2001,G.O2014}.
Because of no any experimental data available for the remaining four
compounds of TiSe, TiTe, HfS, and HfTe, here we have assumed that
they also crystallizes in the same WC-type structure. For five
experimentally known compounds TiS, ZrS, ZrSe, ZrTe and HfSe, our
DFT calculations yield the good agreement of their equilibrium
lattice parameters with the experimental data as shown in Table S1.
Furthermore, their enthalpies of formation are derived in Table S1,
indicating their stabilities in the thermodynamics.

\subsection{Electronic band structures}

To elucidate the electronic band structure of these compounds, we
have first repeated the calculations of ZrTe and obtained the
electronic band structures are very similar to the reported data in
Ref. \onlinecite{H.Weng_12016}, indicating the reliability of our
current calculations. Remarkably, the derived electronic band
structures of other compounds in this family are all similar to that
of ZrTe.

We have elucidated the electronic band structures of these nine
compounds. Interestingly, their electronic structures are in
similarity to the case of ZrTe in Ref. \onlinecite{H.Weng_12016}. We
have selected ZrSe as a prototypical example to show the crucial
feature of electronic structures (details refer to Fig. S1-S3 in
supplementary materials). Without the spin-orbit coupling (SOC)
effect the two main features can be observed: In the first, a Dirac
nodal line (DNL as marked in Fig. \ref{fig1}(c) in the main text)
centered at each $K$ point in the $K_z$ = 0 plane is formed around
the Fermi level due to the linear crossing of the inverted bands
between the Zr $d_{xz}$+$d_{yz}$ orbitals and Zr
d$_{x^2-y^2}$+d$_{xy}$ orbitals (Fig. s1(a)). In the second, a
six-fold degenerate nodal point (Fig. s1(a)) locating at (0, 0,
0.3025) along the $\Gamma$-A direction around the Fermi level due to
another band inversion between the doubly degenerate Zr
$d_{xz}$+$d_{yz}$ and the Zr $d_{z^2}$-like orbitals at the A point
of the BZ. Because the masses of both Zr and Se are not so light
that their SOC effects can not be ignored. With the SOC inclusion,
the derived electronic band structure clearly exhibits the apparent
changes around the Fermi level: Firstly, due to the lack of
inversion symmetry the spin splitting bands appears and each DNL
around the K point is indeed broken into two Weyl points (WPs) with
the opposite chirality (see WP+ and WP- as marked in Fig. s1(c)). In
total, there are six pairs of WPs locating at both $k_z$ =
$\pm$0.01628 plane slightly above and below the $k_z$ = 0 planes.
All these twelve WPs have the same energy level (Fig. s1(d)).
Secondly, the SOC inclusion splits each six-fold generated nodal
point into two triply degenerate nodal points (TDNP1 (0, 0, 0.2904)
and TDNP2 (0, 0, 0.3146) as marked in Fig. s1(c) (their specified
locations are marked in Fig. \ref{fig1}(c) in the main text) along
the $\Gamma$ to $A$ direction. Their appearance is protected by the
C$_{3z}$ rotation and mirror symmetries, being the same as both ZrTe
and TaN cases have \cite{H.Weng_02016,H.Weng_12016}.

The evolution of the derived electronic band structures around one
of WPs for ZrSe is illustrated in supplementary Fig. S1(a,b,c) and
it can be clearly seen that the WP appears around the Fermi level in
supplementary Fig. S2c. Their non-trivial topological property of
the electronic bands can be identified using the Wilson loop method
\cite{add1} (see Fig. S2). As shown in Fig. S2(d and e), the
calculated evolution of the Wannier centers formed along the k$_y$
direction in the two k$_z$ = 0 and $\pi$ planes. It can be seen that
the Z$_2$ numbers (namely, counting the times of Wannier center
crosses a reference line) of both these planes are odd, indicating
their topological non-trivial feature. We have also derived the
topological non-trivial (0001) and (10$\bar{1}$0) Fermi-arc surface
states in Figs. S3 and S4, showing the very similar non-trivial
surface states to ZrTe \cite{H.Weng_12016}. Besides ZrSe and ZrTe,
the other seven members in this family all exhibit the similar
electronic structures featured with the coexisted TDNPs and WPs in
their electronic structures of bulk phases.

\begin{table}[!t]
\begin{center}
\caption{DFT-derived lattice constants $a$ (\AA) and $c$ (\AA) and
enthalpy of formation (eV/atom) of single crystals, in comparison
with available experimental data. \label{tab2}}
\begin{ruledtabular}
\begin{tabular}{ccccc}
                & \emph{a}           & \emph{c}     & $ \Delta$H   & \\
\hline
 TiS  &  3.287    &  3.210            &               &  Expt. Ref. \onlinecite{Hahn_01959}  \\
      &  3.267    &  3.223            &    -1.50   & Calc. \\
 \hline
 TiSe & 3.419    &  3.402                       &    -1.28                  & Calc.   \\
 \hline
 TiTe & 3.669    &  3.656                      &    -0.64                 & Calc.  \\
 \hline
 ZrS  & 3.446    &  3.445                           &                      &    Exp. Ref. \onlinecite{Harry1957,Steiger1970}   \\
      & 3.460    &  3.475                     &    -1.65                  & Calc.  \\
 \hline
 ZrSe$_{0.90}$   & 3.551   &  3.615                           &                      &   Exp. Ref. \onlinecite{Hahn1959}   \\
 ZrSe            & 3.584   &  3.649                    &    -1.49                 & Calc.  \\
 \hline
 ZrTe            &  3.760 & 3.860 & & Exp. Ref.
 \onlinecite{Sodeck1979,G.O2001,G.O2014} \\
                 & 3.798 & 3.895 & -0.91 & Calc.\\
 \hline
 HfS             & 3.395    &  3.447                    &    -1.54                 & Calc.   \\
 \hline
 Hf$_{0.92}$Se   &  3.4958    &  3.6069                           &                      &   Exp. Ref. \onlinecite{Schewe1994}   \\
                 & 3.5173    &  3.6365                   &    -1.32               & Calc.   \\
 \hline
 HfTe            & 3.739    &  3.885                   &    -0.68               & Cal.   \\
\end{tabular}
\end{ruledtabular}
\end{center}
\end{table}

\begin{figure}
\includegraphics[height=0.48\textwidth]{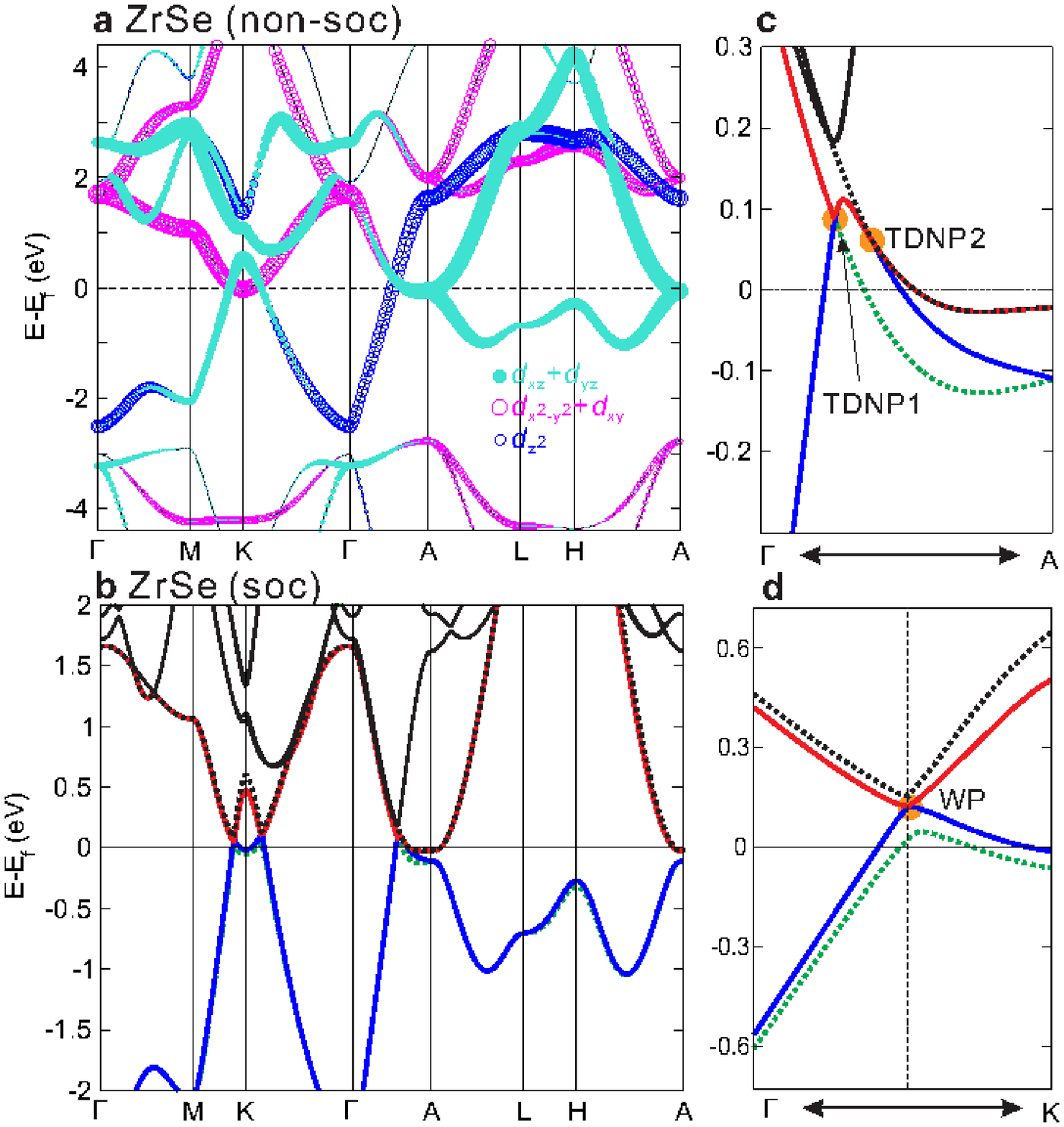}
\caption{Electronic structures of ZrSe. Panel (\textbf{a}): band
structure without the SOC inclusion shows (i) the band inversion
between $d_{xz}$+$d_{yz}$ and $d_{x^2-y^2}$+d$_{xy}$ at K and (ii)
the band inversion between $d_z^2$ and $d_{xz}$+$d_{yz}$ orbitals at
A. Panel (\textbf{b}): Electronic band structure with the SOC
inclusion. Panel (\textbf{c}) shows the zoom-in visualization of two
TDNP1 and TDNP2 along the $\Gamma$-A direction in panel (b), whereas
Panel (\textbf{d}) shows the zoom-in bands crossing one WP (0.27314,
0.27314, $\pm$0.01628) around the Fermi level with the SOC
inclusion.}
\end{figure}

\begin{figure}
\includegraphics[height=0.35\textwidth]{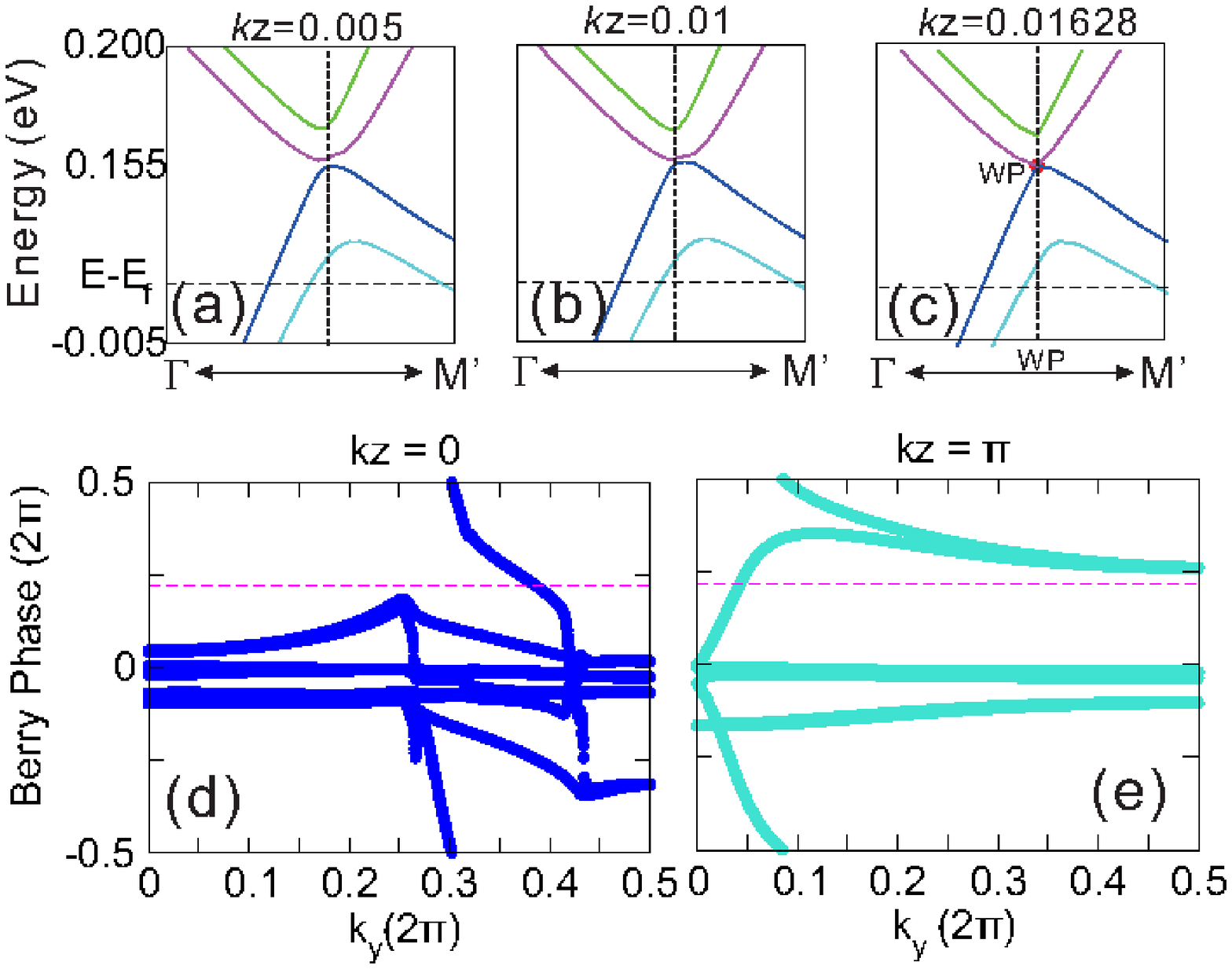}
\caption{Electronic band structures around the WP node at $K_z$ =
0.005 in panel (a), 0.010 in panel (b), and 0.01628 (exactly
corresponding to the WP node) in panel (c), respectively. Panels (d
and e) denote the derived Wilson loops of ZrSe which show the $k_y$
evaluation of the Berry phases of all occupied bands along the $k_x$
direction in both $k_z$ = 0 and $k_z$ = $\pi$ planes, respectively.
} \label{fig3s}
\end{figure}

To further clarify the topological feature in ZrSe, we have
calculated the surface electronic structures on its (0001) and
(10$\bar{1}$0) surfaces (see supplementary Fig. S2), clearly
indicating the topological surface states. In addition, we also plot
the Fermi surface of both surfaces in supplementary Fig. S3. On the
(0001) Fermi surface the two TDNPs along $\Gamma$ to $A$ direction
in its bulk phase are both projected onto the $\bar{\Gamma}$ point,
becoming invisible due to the overlapping with the projection of the
bulk electronic bands. Each pair of WPs above and below the $k_z$ =
0 plane which have different chirality in the bulk phase will be
projected onto the same point, totally forming six projected nodes.
These six projected nodes on the (0001) surface are further
connected by Fermi arcs, resulting in the appearance of two
triangle-like loops, as illustrated in supplementary Fig. S3(a). It
is impossible to see these six projected nodes in supplementary Fig.
S3(a) because their energy is 155 meV above the Fermi level. By
changing the chemical potential to 155 meV, the Fermi surface gives
rise to the clear visualization of six projected nodes in
supplementary Fig. S3(c). In order to visualize the Weyl nodes, we
further calculated (10$\bar{1}$0) Fermi surface at the energy level
of 155 meV above the Fermi level in supplementary Fig. S3(d). On
this surface, the six pairs of WPs with opposite chirality are
projected to different positions. Two WPs with same chirality are
projected to the same point on the (10$\bar{1}$0) surface (called
WP1) and the projected points of other WPs are labeled as WP2. It
can be clearly seen that the projected Weyl points are connected by
Fermi arcs. For each WP1 point, there is one arc connecting it by
going through the $\bar{\Gamma}$ - $\bar{M}$ path, whereas, for each
WP2 there are two arcs connecting them in supplementary Fig. S3(d).
In addition, on both (0001) and (10$\bar{1}$0) surfaces it is
impossible to see TDNPs because the projection of TDNPs are all
overlapped with bulk electronic bands, as illustrated in Fig.
supplementary Fig. S3(c).

\begin{figure*}
\includegraphics[height=0.35\textwidth]{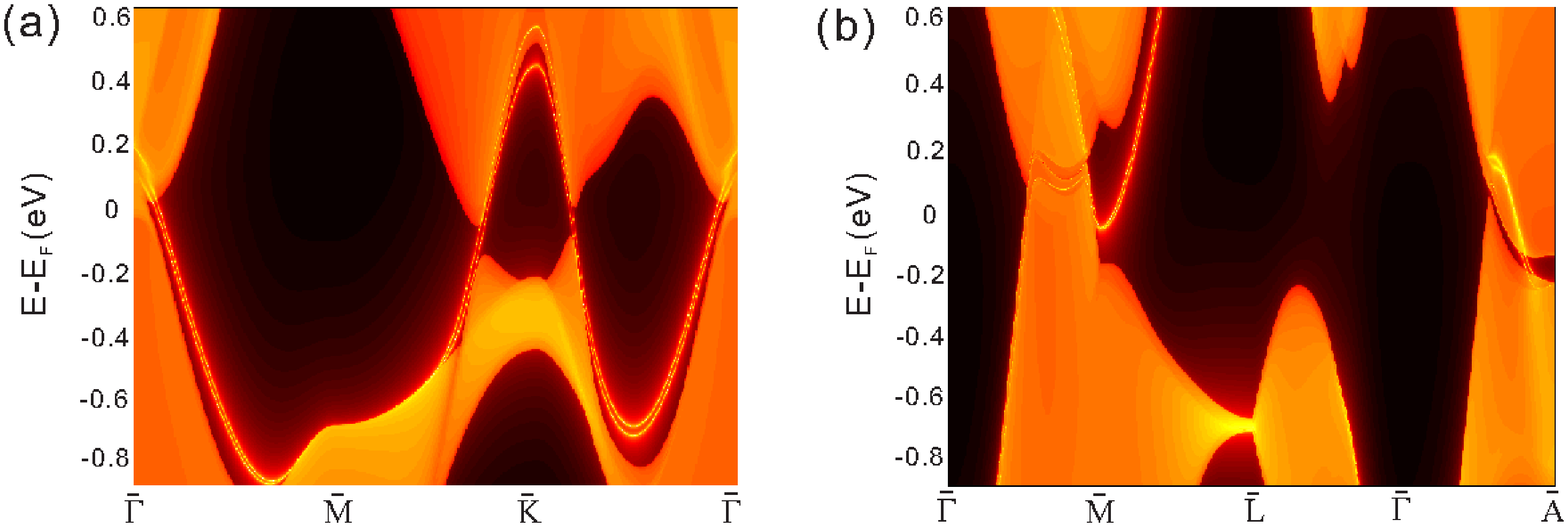}
\caption{Calculated surface electron structures of both (0001) and
(10$\bar{1}$0) surfaces of ZrSe.} \label{fig4s}
\end{figure*}

\begin{figure*}
\includegraphics[height=0.80\textwidth]{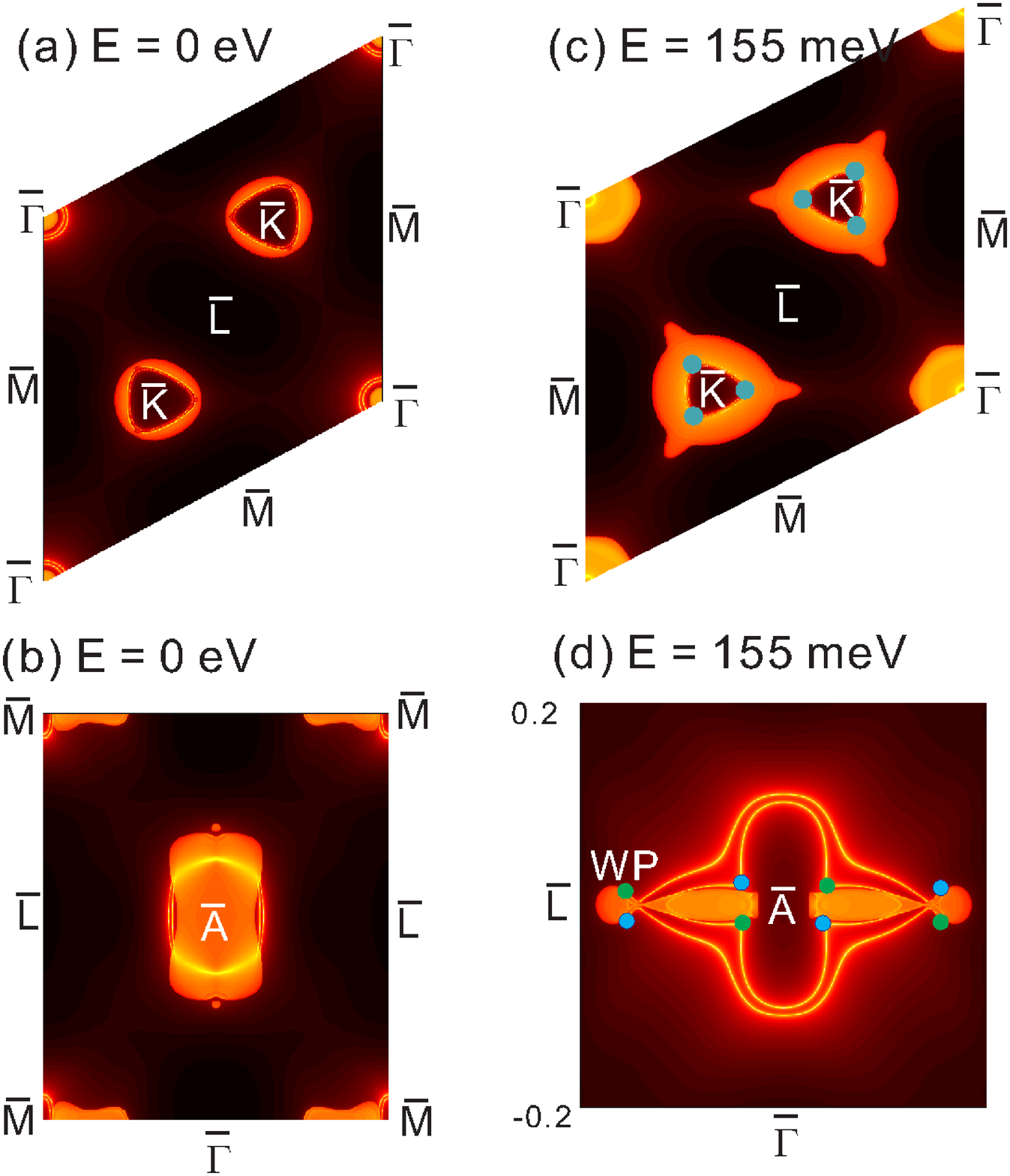}
\caption{The Fermi surfaces of (0001) (panels: a and c) and
(10$\bar{1}$0) (panels: b and d) surfaces of ZrSe. Panels (a) and
(b) denote the Fermi surfaces at the Fermi level, while panels (c)
and (d) refer to the Fermi surfaces with a chemical potential of 155
meV above the Fermi level. The solid circles indicate the projected
WPs in the (10$\bar{1}$0) surfaces.} \label{fig5s}
\end{figure*}

Through our calculations the other eight members in this family all
exhibit the similar electronic structures with the coexisted TDNPs
and WPs in their bulk phases, except for TiS. Because of the weak
SOC effect, TiS is highly unique with the coexisted six Dirac nodal
lines (DNLs) and two six-degenerated nodal points, which is exactly
what happened for ZrSe without the SOC effect.

\subsection{Phonon dispersions of MX}

Supplementary Figure S4 compiles the DFT-derived phonon dispersions
of all nine $MX$ compounds. Among them, only three compounds of TiS,
ZrSe and HfTe exhibit the non-trivial topological phonon states with
the appearance of the triply degenerate nodal points (TDNPs as
marked in supplementary Figure S4, which refers to three-component
bosons.

\begin{figure*}
\includegraphics[height=0.80\textwidth]{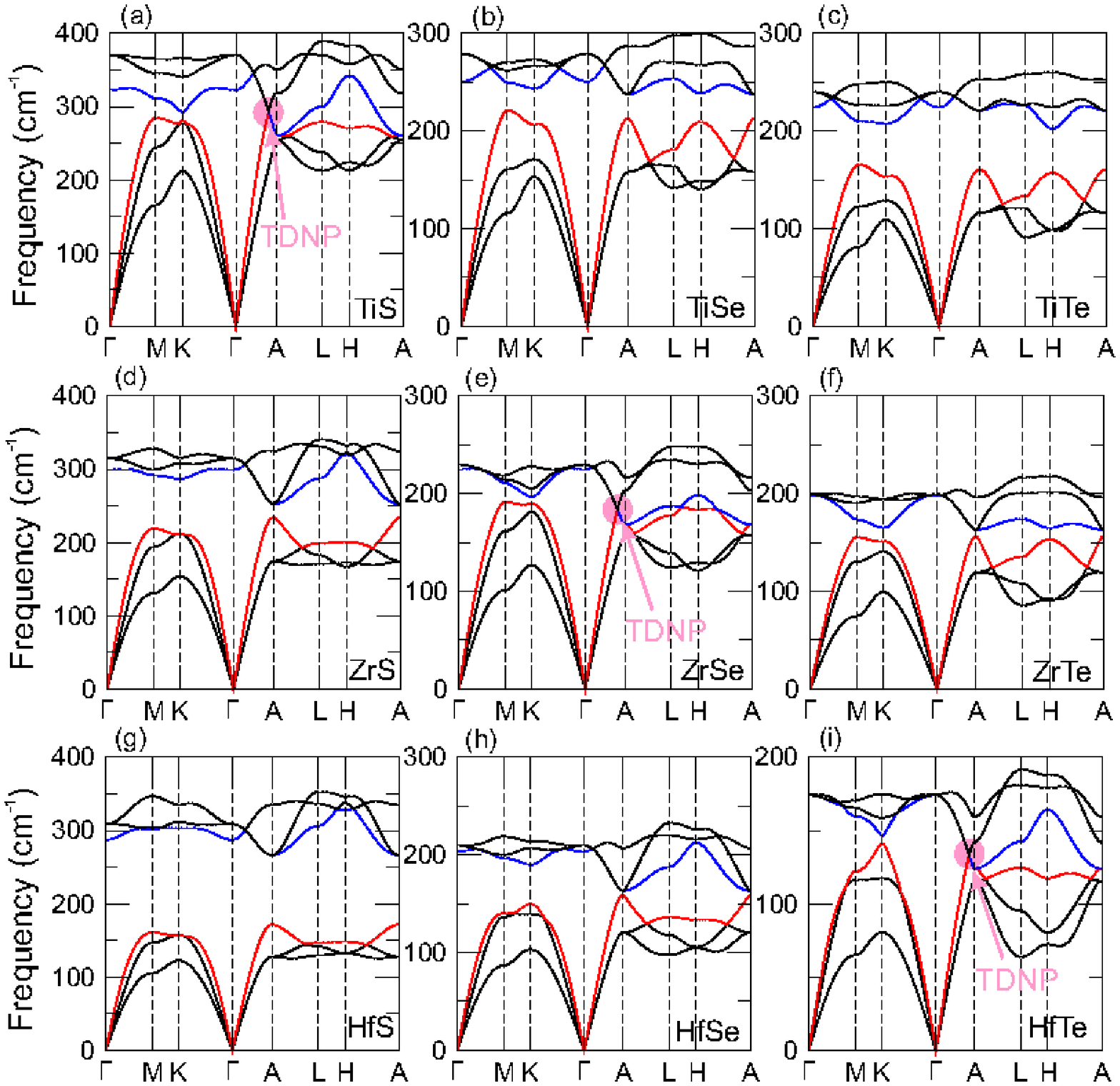}
\caption{DFT-derived phonon dispersions of all nine \emph{MX}
compounds with the WC-type structure at their optimized equilibrium
lattices.} \label{fig5s}
\end{figure*}


\begin{thebibliography}{18}

\bibitem{H.Weng2016} Weng, H.M., Dai, X., \& Fang, Z. Topological semimetals
predicted from first-principles calculations, \emph{J. Phys. Condens.
Matter.} \textbf{28},303001 (2016).

\bibitem{Y.B2016} Yan, B.H. \& Felser, C. Topological Materials:
Weyl semimetals, Annu. Rev. \emph{Condens. Matter Phys.} \textbf{8}, 337--354
(2016).

\bibitem{C.K2016} Chiu, C.-K., Teo, J. C. Y., Schnyder, A. P.
\& Ryu, S.  Classification of topological quantum matter with
symmetries,\emph{ Rev. Mod. Phys.} \textbf{88}, 035005 (2016).

\bibitem{A.Bansil2016}Bansil, A., Lin, H., \& Das, T.  Colloquium:
Topological band theory, \emph{Rev. Mod. Phys.} \textbf{88}, 021004 (2016).

\bibitem{S.Rao} Rao, S. Weyl semi-metals: A short review, arXiv:1603.02821 (2016).


\bibitem{Z.Wang2012} Wang, Z. J., Sun, Y., Chen, X.-Q., Franchini, C., Xu,
G., Weng,H. M., Dai, X. \& Fang, Z.
Dirac semimetal and topological phase transitions in
A$_{3}$Bi (A = Na, K, Rb), \emph{Phys. Rev. B} \textbf{85}, 195320 (2012).

\bibitem{Young2012}Young, S. M., Zaheer, S., Teo, J. C. Y., Kane, C. L.,
Mele, E. J.\& Rappe, A. M. Dirac Semimetal in Three Dimension. \emph{Phys.
Rev. Lett.} \textbf{108}, 140405 (2012).

\bibitem{Z.W2013}Wang,Z. ,Weng, H. , Wu, Q.,Dai, X.  \& Fang, Z.
Three dimensional Dirac semimetal and quantum transport in
Cd$_3$As$_2$, \emph{Phys. Rev. B} \textbf{88}, 125427 (2013).

\bibitem{Cheng.X2014}Cheng, X. Y., Li, R. H., Sun, Y., Chen, X.-Q.,
Li, D. Z. \& Li, Y. Y. Ground-state phase in the three-dimensional
topological Dirac semimetal Na$_3$Bi. \emph{Phys. Rev. B} \textbf{89}, 245201
(2014).

\bibitem{Neupane2014}Neupane, M., Xu, S. Y., Sankar, R., Alidoust, N.,
Bian, G., Liu, C., Belopolski, I., Chang, T. R., Jeng, H. T., Lin,
H., Bansil, A., Chou, F. \& Hasan, M. Z. Observation of a
three-dimensional topological Dirac semimetal phase in high-mobility
Cd$_3$As$_2$. \emph{Nat. Commun.} \textbf{5}, 3786 (2014).

\bibitem{Z.K2014}Liu, Z. K., Jiang, J., Zhou, B., Wang, Z.J., Zhang, Y., Weng, H. M.,
Prabhakaran, D., Mo, S.-K., Peng, H., Dudin, P., Kim, T., Hoesch,
M., Fang, Z., Dai, X., Shen, Z. X., Feng, D. L., Hussain, Z., \&
Chen, Y. L. A stable three-dimensional topological Dirac semimetal
Cd$_3$As$_2$. \emph{Nature Mater.} \textbf{13}, 677 (2014).

\bibitem{Liu.Z2014}Liu, Z. K., Zhou, B., Zhang, Y., Wang, Z. J., Weng, H. M.,
Prabhakaran, D., Mo, S.-K., Shen, Z. X., Fang, Z., Dai, X.,
Hussain,Z. \& Chen, Y. L. Discovery of a Three-Dimensional
Topological Dirac Semimetal, Na$_3$Bi. \emph{Science} \textbf{343}, 864-867 (2014).

\bibitem{B.J2014} Yang, B. J. \& Nagaosa, N.  Classification of
stable three dimensional Dirac semimetals with nontrivial topology,
\emph{Nat. Commun.} \textbf{5}, 4898 (2014).

\bibitem{Xu2015}Xu, S.-Y., Liu, C., Kushwaha, S. K., Sankar, R., Krizan, J. W.,
Belopolski, I., Neupane, M., Bian, G., Alidoust, N., Chang, T. R.,
Jeng, H. T., Huang, C. Y., Tsai, W. F., Lin, H., Shibayev, P.
P.,Chou, F. C., Cava, R. J. \& Hasan. M. Z. Observation of Fermi arc
surface states in a topological metal. \emph{Science} \textbf{347}, 294-298 (2015).

\bibitem{Du.Y2015} Du, Y., Wan, B., Wang, D., Sheng, L., Duan, C.G., \& Wan,
X.G. Dirac and Weyl Semimetal in XYBi (X= Ba, Eu; Y= Cu, Ag and Au).
\emph{Sci. Rep.} \textbf{5}, 14423 (2015).

\bibitem{J.Hul} Hu, J., Zhu, Y.L., Graf, D., Tang, Z.J., Liu, J.Y., \& Mao, Z.Q.
Quantum oscillation studies of topological semimetal candidate ZrGeM
(M=S, Se, Te), \emph{Phys. Rev. B} \textbf{95},205134 (2017).

\bibitem{S.Murakami2007} Murakami, S. Phase transition between the
quantum spin Hall and insulator phases in 3d: Emergence of a
topological gapless phase, \emph{New J. Phys.} \textbf{9}, 356 (2007).

\bibitem{X.Wan2011} Wan, X.G., Turner, A. M., Vishwanath, A. \& Savrasov, S. Y.
Topological semimetal and Fermi-arc surface states in the electronic
structure of pyrochlore iridates, \emph{Phys. Rev. B} \textbf{83},205101 (2011).

\bibitem{G.Xu2011} Xu, G., Weng H., Wang,Z.,Dai, X.\& Fang, Z. Chern
Semimetal and the Quantized Anomalous Hall Effect in HgCr$_2$Se$_4$,
\emph{Phys. Rev. Lett.} \textbf{107}, 186806 (2011).

\bibitem{S.Y2015} Xu, S.-Y., Alidoust, N., Belopolski, I., Yuan, Z.,
Bian, G., Chang, T.-R., Zheng, H., Strocov, V. N., Sanchez, D. S.,
Chang, G., Zhang, C., Mou, D., Wu, Y., Huang, L., Lee, C.-C., Huang,
S.-M., Wang, B., Bansil, A., Jeng, H.-T., Neupert, T., Kaminski, A.,
Lin, H., Jia, S. \& Hasan, M. Z. Discovery of a Weyl fermion state
with Fermi arcs in niobium arsenide. \emph{Nature Phys.} \textbf{11}, 748-754
(2015).

\bibitem {Shekhar2015}Shekhar, C., Nayak, A. K., Sun, Y., Schmidt, M.,
Nicklas, M., Leermakers, I., Zeitler, U., Skourski,Y., Wosnitza, J.,
Liu, Z.K., Chen,Y.L., Schnelle, W., Borrmann,H., Grin, Y., \&
Felser, C. \& Yan, B.H. Extremely large magnetoresistanceand
ultrahigh mobility in the topological Weyl semimetal candidate NbP.
\emph{Nature Phys.} \textbf{11}, 645-649 (2015)

\bibitem{S.Y_02015} Xu, S.-Y., Belopolski, I., Sanchez, D. S., Guo, C.,
Chang, G., Zhang, C., Bian, G., Yuan, Z., Lu, H., Feng, Y., Chang,
T.-R., Shibayev, P. P., Prokopovych, M. L., Alidoust N., Zheng, H.,
Lee, C.-C., Huang, S.-M., Sankar, R., Chou, F., Hsu,C.-H., Jeng,
H.-T., Bansil, A., Neupert,T., Strocov, V. N., Lin, H., Jia, S. \&
Hasan, M. Z. Experimental discovery of a topological Weyl semimetal
state in TaP. \emph{Sci. Adv.} \textbf{1}, e1501092 (2015).

\bibitem{H.Weng2015} Weng, H.M., Fang, C., Fang, Z., Bernevig, B. A.
\& Dai,X. Weyl Semimetal Phase in Noncentrosymmetric
Transition-Metal Monophosphides, \emph{Phys. Rev. X} \textbf{5}, 011029 (2015).

\bibitem{S.M2014} Huang, S.M., Xu, S.Y., Belopolski, I., Lee, C.C.,
Chang, G., Wang, B.K., Alidoust, N., Bian, G., Neupane, M., Zhang,
C., Jia, S., Bansil, A., Lin, H.\& Hasan, M.Z. AWeyl Fermion
semimetal with surface Fermi arcs in the transition metal
monopnictide TaAs class, \emph{Nat. Commun.} \textbf{6}, 7373 (2015).

\bibitem{B.Q2015} Lv, B.Q., Weng, H. M., Fu, B.B., Wang, X. P., Miao, H.,
Ma, J., Richard, P., Huang, X. C., Zhao, L. X., Chen, G. F., Fang,
Z., Dai, X., Qian, T. \& Ding, H. Experimental Discovery of Weyl
Semimetal TaAs, \emph{Phys. Rev. X} \textbf{5}, 031013 (2015).

\bibitem{B.Q_02015}Lv, B.Q., Xu, N., Weng, H. M., Ma, J. Z., Richard, P.,
Huang, X.C., Zhao, L.X., Chen, G.F., Matt, C.E., Bisti, F., Strocov
V.N., Mesot, J., Fang, Z., Dai, X., Qian, T., Shi, M.  \& Ding, H.
Observation of Weyl nodes in TaAs, \emph{Nature Phys.} \textbf{11}, 724-728 (2015).

\bibitem{S.-Y2015}Xu, S.-Y., Belopolski, I., Alidoust, N., Neupane, M.,
Bian, G., Zhang, C., Sankar, R., Chang, G., Yuan, Z., Lee, C.-C.,
Huang, S.-M., Zheng, H., Ma, J., Sanchez, D. S., Wang, B. , Bansil,
A., ChouF., Shibayev, P.P., Lin, H., Jia, S. \& Hasan, M. Z.
Discovery of a Weyl fermion semimetal and topological Fermi arcs,
\emph{Science} \textbf{349}, 613-617 (2015).

\bibitem{L.Yang2015} Yang, L., Liu, Z., Sun, Y., Peng, H., Yang, H.,
Zhang, T., Zhou, B., Zhang, Y., Guo, Y., Rahn, M., Prabhakaran, D.,
Hussain, Z., Mo, S.-K., Felser, C., Yan, B.H. \& Chen, Y.L.  Weyl
semimetal phase in the non-centrosymmetric compound TaAs, \emph{Nature Phys.}
\textbf{11}, 728-734 (2015).

\bibitem{Y.Zhang2016} Zhang, Y., Bulmash, D., Hosur, P., Potter, A. C.
\& Vishwanath, A.  Quantum oscillations from generic surface Fermi
arcs and bulk chiral modes in Weyl semimetals, \emph{Sci. Rep.} \textbf{6}, 23741
(2016).

\bibitem{A.A2015}Soluyanov, A. A, Gresch,D ., Wang, Z.J., Wu, Q.,Troyer, M.,
Dai, X. \& Bernevig, B A. Type-II Weyl semimetals, \emph{Nature} \textbf{527}, 495-498 (2015).

\bibitem{Xu SY2016} Xu, S.Y., Alidoust, N., Chang, G., Lu, H., Singh, B.,
Belopolski, I., Sanchez, D., Zhang, X., Bian, G., Zheng, H., \&
Husanu, M.A. Discovery of Lorentz-violating Weyl fermion semimetal
state in LaAlGe materials. arXiv preprint arXiv:1603.07318, (2016).

\bibitem{Chang2016}Chang, G., Singh, B., Xu, S.Y., Bian, G., Huang,
S.M., Hsu, CH., Belopolski, I.,
Alidoust, N., Sanchez, D.S., Zheng, H., \& Lu, H. Magnetic and
noncentrosymmetric Weyl fermion semimetals in the RAlX family of
compounds (R= rare earth, Al= aluminium, X= Si, Ge). arXiv preprint
arXiv:1604.02124, (2016).

\bibitem{Yang2016}Yang,H., Sun,Y., Zhang,Y., Shi,W.J., Parkin,S.S., \& Yan,
B.H. Topological Weyl semimetals in the chiral antiferromagnetic
materials Mn$_3$Ge and Mn$_3$Sn. \emph{New J. Phys.} \textbf{19}, 015008 (2017).

\bibitem{Singh2012}Singh, B., Sharma,A., Lin, H., Hasan, M.Z., Prasad, R.,
\& Bansil, A. Topological electronic structure and Weyl semimetal in
the TlBiSe$_2$ class of semiconductors. \emph{Phys. Rev. B.} \textbf{86}, 115208 (2012).

\bibitem{Ruan2016}Ruan, J., Jian, S.K., Yao, H., Zhang, H., Zhang, S.C.\& Xing,
D. Symmetry-protected ideal Weyl semimetal in HgTe-class materials.
\emph{Nat. Commun.} \textbf{7}, 11136 (2016).

\bibitem{D.T2013}Son, D.T., \& Spivak, B. Z. Chiral anomaly and classical
negative magnetoresistance of Weyl metals, \emph{Phys. Rev. B} \textbf{88}, 104412
(2013).

\bibitem{X.Huang2015}Huang, X., Zhao, L., Long, Y., Wang, P., Chen, D.,
Yang, Z., Liang, H., Xue, M., Weng, H., Fang, Z., Dai, X. \& Chen,
G. Observation of the Chiral-Anomaly-Induced Negative
Magnetoresistance in 3D Weyl Semimetal TaAs, \emph{Phys. Rev. X} \textbf{5}, 031023
(2015).

\bibitem{P.Hosur2013}Hosur, P.\& Qi, X. Recent developments in
transport phenomena in Weyl semimetals, \emph{C. R. Phys.} \textbf{14}, 857-870 (2013).

\bibitem{Fang.C2016}Fang, C., Weng, H. M., Dai, X. \& Fang,
Z. Topological nodal line semimetals. \emph{Chin. Phys. B} \textbf{25}, 117106
(2016).

\bibitem{Ryu2002}Ryu, S. \& Hatsugai, Y. Topological Origin of
Zero-Energy Edge States in Particle-Hole Symmetric Systems. \emph{Phys.
Rev. Lett.} \textbf{89}, 077002 (2002).

\bibitem{Heikkila2011} Heikkil\"{a}, T. T. \& Volovik, G. E. Dimensional
crossover in topological matter: Evolution of the multiple Dirac
point in the layered system to the flat band on the surface. \emph{JETP
Lett.} \textbf{93}, 59-65 (2011).

\bibitem{Burkov.A2011} Burkov, A. A., Hook, M. D. \& Balents, L.
Topological nodal semimetals. \emph{Phys. Rev. B} \textbf{84}, 235126 (2011).

\bibitem{Ronghan2016}Li, R. H., Ma, H., Cheng, X. Y., Wang, S. L.,
Li, D. Z., Zhang, Z.Y., Li, Y. Y. \& Chen, X.-Q. Dirac node lines in
pure alkali earth metals. \emph{Phys. Rev. Lett.} \textbf{117}, 096401 (2016).

\bibitem{Weng.H.M2015}Weng, H. M., Liang, Y. Y., Xu, Q. N., Yu, R., Fang,
Z., Dai,X. \& Kawazoe, Y. Topological node-line semimetal in three
dimensional graphene networks. \emph{Phys. Rev. B} \textbf{92}, 045108 (2015).

\bibitem{Yu.R2015} Yu, R., Weng, H. M., Fang, Z., Dai, X. \& Hu, X.
Topological Node-Line Semimetal and Dirac Semimetal State in
Antiperovskite.Cu$_3$PdN. \emph{Phys. Rev. Lett.} \textbf{115}, 036807 (2015).

\bibitem{Kim2015}Kim, Y., Wieder, B. J., Kane, C. L. \& Rappe, A. M.
Dirac Line Nodes in Inversion-Symmetric Crystals. \emph{Phys. Rev. Lett.}
\textbf{115}, 036806 (2015).

\bibitem{Xie.L2015}Xie, L. S., Schoop, L. M., Seibel, E. M., Gibson,
 Q. D., Xie, W. W. \& Cava, R. J. A new form of Ca$_3$P$_2$ with a ring of
 Dirac nodes. \emph{APL Materials} \textbf{3}, 083602 (2015).

\bibitem{M.G.Zeng2015} Zeng, M. G., Fang, C., Chang, G. Q., Chen, Y.-A., Hsieh, T.,
Bansil, A., Lin, H. \& Fu, L. Topological semimetals and topological
insulators in rare earth monopnictides. Preprint at
https://arxiv.org/abs/1504.03492 (2015).

\bibitem{Mullen2015}Mullen, K., Uchoa, B. \& Glatzhofer, D. T. Line of Dirac
Nodes in Hyperhoneycomb Lattices. \emph{Phys. Rev. Lett.} \textbf{115}, 026403
(2015).

\bibitem{Gan2016}Gan, L.-Y., Wang, R., Jin, Y. J., Ling, D. B., Zhao, J. Z.,
Xu, W.P., Liu, J. F. \& Xu, H. Pressure-induced Topological
Node-Line Semimetals in Alkaline-Earth Hexaborides XB$_6$ (X=Ca, Sr,
Ba). Preprint at https://arxiv.org/abs/1611.06386 (2016).

\bibitem{Kawakami2016}Kawakami, T. \& Hu, X. Symmetry-Guaranteed
and Accidental Nodal-Line Semimetals in FCC Lattice. Preprint at
https://arxiv.org/abs/1611.07342 (2016).

\bibitem{Yang2017}Yang, B., Zhou, H. C., Zhang, X. M., Liu, X. B. \& Zhao,
M. W. Dirac cones and highly anisotropic electronic structure of
supergraphyne. \emph{Carbon} \textbf{113}, 40-45 (2017).

\bibitem{Li2017} Li, J.X., Ma, H., Feng, S., Ullah, S., Li, R., Dong, J.
\& Chen, X.-Q. Topological nodal line states and a potential
catalyst of hydrogen evolution in the TiSi family. arXiv preprint
arXiv:1704.07043 (2017).

\bibitem{B.Bradlyn2016}Bradlyn,B., Cano,J., Wang, Z., Vergniory, M. G.,
Felser, C., Cava, R. J. \&  Bernevig, B. A. Beyond Dirac and Weyl
fermions: Unconventional quasiparticles in conventional crystals,
\emph{Science} \textbf{353}, 558 (2016).

\bibitem{G.W2016}Winkler, G. W., Wu, Q., Troyer, M., Krogstrup, P.\&
Soluyanov, A. A. Topological Phases in InAs$_{1-x}$Sb$_{x}$ : From
Novel Topological Semimetal to Majorana Wire, \emph{Phys. Rev. Lett.} \textbf{117}, 076403 (2016).

\bibitem{H.Weng_02016}Weng, H., Fang, C., Fang, Z.\& Dai, X.  Topological
semimetals with triply degenerate nodal points in $\theta$-phase
tantalum nitride, \emph{Phys. Rev. B} \textbf{93}, 241202 (2016).

\bibitem{H.Weng_12016}Weng, H., Fang, C., Fang, Z. \& Dai, X.  Coexistence
of Weyl fermion and massless triply degenerate nodal points, \emph{Phys.
Rev. B} \textbf{94}, 165201 (2016).

\bibitem{Zhu2016}Zhu, Z., Winkler, G. W. ,Wu, Q. S., Li, J. \& Soluyanov, A. A.
Triple point topological metals. \emph{Phys. Rev. X} \textbf{6}, 031003 (2016).

\bibitem{G.Chang2016} Chang, G., Xu, S.Y., Huang, S.M., Sanchez, D.S.,
Hsu, C.H, Bian, G., Yu, Z.M., Belopolski, I., Alidoust, N., Zheng,
H.,Chang, T.R., Jeng, H.J., Yang, S.A., Neupert, T., Lin, H. \&
Hasan, M.Z. New fermions on the line in topological symmorphic
metals. \emph{Sci. Rep.} \textbf{7}, 1688 (2017).

\bibitem{He.J2017} He, J. B., Chen, D., Zhu, W. L.,Zhang, S. ,Zhao, L. X.,
Ren, Z. A.\& Chen, G. F. . Magnetotransport properties of the triply
degenerate node topological semimetal: tungsten carbide. \emph{Phys. Rev. B}
\textbf{95}, 195165 (2017)

\bibitem{Ding.H2017} Lv, B.Q., Feng, Z.L., Xu, Q.N., Gao, X., Ma,
J.Z., Kong, L.K., Richard, P., Huang, Y.B., Strocov, V. N., Fang,
C., Weng, H.M., Shi, Y.G., Qian, T., \& Ding, H. Observation of
three-component fermions in the topological semimetal molybdenum
phosphide, \emph{Nature} \textbf{546}, 627-631 (2017).

\bibitem{H.Yang2017} Yang, H., Yu, J.B., Parkin, S.S.P., Felser, C.,
Liu, C.-X. \& Yan, B.H., Prediction of triple point fermions in
simple half-Heusler topological insulators, \emph{Phys. Rev. Lett.} \textbf{119},
136401 (2016).

\bibitem{J.Yu2017} Yu, J.B., Yan, B.H. \& Liu, C.-X. Model
Hamiltonian and time reversal breaking topological phases of
anti-ferromagnetic half-Heusler Materials, \emph{Phys. Rev. B}
\textbf{95}, 235158 (2017)

\bibitem{Lu.L2013}Lu, L., Fu, L., Joannopoulos, J. D. \&
Solja$\breve{c}$i$\acute{c}$, M. Weyl points and line nodes in
gyroid photonic crystals. \emph{Nature Photon.} \textbf{7}, 294-299 (2013).

\bibitem{Lu.L2015}Lu, L., Wang, Z., Ye, D., Ran, L., Fu, L., Joannopoulos, J. D.
\& Soljacic, M. Experimental observation of Weyl points, \emph{Science}
\textbf{349}, 622 (2015).

\bibitem{Huber.S2016} Huber, S.D. Topological mechanics. \emph{Nature Phys.} \textbf{12}, 621-623 (2016).

\bibitem{Prodan.E2009} Prodan, E. \& Prodan, C.
Topological phonon modes and their role in dynamic instability of
microtubules, \emph{Phys. Rev. Lett.} \textbf{103}, 248101 (2009).

\bibitem{Chen.B2014} Chen, B.G.G., Upadhyaya, N. \& Vitelli, V.
Nonlinear conduction via solitons in a topological mechanical insulator.
\emph{Proc Natl Acad Sci.} \textbf{111}, 13004-13009 (2014).

\bibitem{Yang.Z2015} Yang, Z., Gao, F., Shi, X., Lin, X. , Gao, Z., Chong, Y.\&
Zhang, B. Topological acoustics. \emph{Phys. Rev. Lett.} \textbf{114}, 114301
(2015).

\bibitem{Wang.P2015}Wang, P., Lu, L. \& Bertoldi, K.
Topological phononic crystals with one-way elastic edge waves. \emph{Phys.
Rev. Lett.} \textbf{115}, 104302 (2015).

\bibitem{Xiao.M2015} Xiao, M., Chen, W.J., He, W.Y.\&
Chan, C. T. Synthetic gauge flux and Weyl points in acoustic
systems. \emph{Nature Phys.} \textbf{11}, 920 (2015).

\bibitem{Nash.L2015}Nash, L. M., Kleckner, D., Read, A. , Vitelli, V., Turner, A. M.
\& Irvine, W. T. Topological mechanics of gyroscopic metamaterials.
\emph{Proc Natl Acad Sci.} \textbf{112}, 14495-14500 (2015).

\bibitem{Susstrunk2015}Susstrunk, R. \& Huber, S. D. Observation of phononic
helical edge states in a mechanical topological insulator, \emph{Science}
\textbf{349}, 47-50 (2015).

\bibitem{Mousavi2015} Mousavi, S. H., Khanikaev, A. B. \& Wang, Z.
Topologically protected elastic waves in phononic metamaterials. \emph{Nat. Commun.} \textbf{6}, 8682 (2015).

\bibitem{Fleury2016}Fleury, R., Khanikaev, A. B. \&  AlRu, A.
Floquet topological insulators for sound, \emph{Nat. Commun.} \textbf{7}, 11744
(2016).

\bibitem{Rocklin2016}Rocklin, D. Z., Chen, B.G.G., Falk, M., Vitelli, V. \& Lubensky,T.
Mechanical weyl modes in topological maxwell lattices. \emph{Phys. Rev.
Lett.} \textbf{116}, 135503 (2016).

\bibitem{He2016}He, C., Ni, X., Ge, H., Sun, X.C., Chen, Y.B., Lu, M.-H., Liu, X.P.
\& Chen, Y.-F. Acoustic topological insulator and robust one-way
sound transport. \emph{Nature Phys.} \textbf{12}, 1124--1129 (2016).

\bibitem{Susstrunk2016}Susstrunk, R. \& Huber, S. D.
Classification of topological phonons in linear mechanical
metamaterials. \emph{Proc Natl Acad Sci.} \textbf{113}, E4767-E4775 (2016).

\bibitem{Lifeng2017} Li, F., Huang, X.Q., Lu, J.Y., Ma, J.H., \&
Liu, Z.Y., Weyl points and Fermi arcs in a chiral phononic crystal,
Nat. Phys. DOI: 10.1038/NPYS42675 (2017).

\bibitem{Zhang.T2017} Zhang, T.T., Song, Z.D., Alexandradinata, A.,
Weng, H.M., Fang, C., Lu, L.\& Fang, Z. Double-Weyl phonons in
traisition-metal monosilicides, arXiv: 1705.07244 (2017).

\bibitem{Zhanglifa2015} Zhang, L.F. \& Niu, Q., Chiral phonons at
high-symmetry points in monolayer hexagonal lattices, \emph{Phys. Rev.
Lett.} \textbf{115}, 115502 (2015).

\bibitem{Liuduan2016} Liu, Y., Xu, Y., Zhang, S.-C., \& Duan W. H.,
Model for topological phononics and phonon diode, \emph{Phys. Rev. B} \textbf{96},
064106 (2017).

\bibitem{Hohenberg.P1964} Hohenberg, P. \& Kohn, W. Inhomogeneous Electron Gas, \emph{Phys. Rev.} \textbf{136}, B864-B871 (1964).

\bibitem{Kohn.W1965} W. Kohn, and L. J. Sham, Self-Consistent Equations Including
Exchange and Correlation Effects, \emph{Phys. Rev.} \textbf{140},
A1133 (1965).

\bibitem{Baroni.S2001} Baroni, S., Gironcoli, S.D., Corso, A.D. \&
Giannozzi, P. Phonons and related crystal properties from
density-functional perturbation theory, \emph{Rev. Mod. Phys.}
\textbf{73}, 515 (2001).

\bibitem{G.Kresse1993} Kresse, G. \& Hafner, J.
Ab initio molecular dynamics for liquid metals. \emph{Phys. Rev. B}
\textbf{47}, 558-561 (1993).

\bibitem{G.Kresse1994}Kresse, G. \& Hafner, J.
Ab initio molecular-dynamics simulation of the liquid-metal
amorphous-semiconductor transition in germanium. \emph{Phys. Rev. B}
\textbf{49}, 14251-14269 (1994).

\bibitem{G.Kresse1996}Kresse, G. \& Furthm\"uller, J. Efficiency of ab-initio total
energy calculations for metals and semiconductors using a plane-wave
basis set. \emph{Comput. Mater. Sci} \textbf{6}, 15-50 (1996).

\bibitem{J.P1992}Perdew, J. P.\& Wang, Y. Accurate and simple analytic
representation of the electron-gas correlation energy, \emph{Phys.
Rev. B} \textbf{45}, 13244-13249 (1992).

\bibitem{J.P1996}Perdew, J. P., Burke, K.\& Ernzerhof, M.
Generalized Gradient Approximation Made Simple, \emph{Phys. Rev.
Lett.} \textbf{77}, 3865-3868 (1996).

\bibitem{P.E1994} Bl\"ohl, P. E. Projector augmented-wave method,
\emph{Phys. Rev. B} \textbf{50}, 17953-17979 (1994).

\bibitem{G.Kresse1999}Kresse, G. \& Joubert, D. From ultrasoft
pseudopotentials to the projector augmented-wave method \emph{Phys.
Rev. B} \textbf{59}, 1758-1775 (1999).

\bibitem{M.P1985} Sancho, M. P., Sancho, J. M. \& Rubio, J. Highly convergent
schemes for the calculation of bulk and surface Green functions,
\emph{J. Phys. F: Met. Phys.} \textbf{15}, 851-858 (1985).

\bibitem{Weng2014} Weng, H.M., Dai, X. \& Fang, Z.  Exploration and prediction
of topological electronic materials based on first-principles
calculations,\emph{ MRS Bull.} \textbf{39}, 849-858 (2014).

\bibitem{Weng2015} Weng, H.M., Yu, R., Hu, X., Dai, X. \& Fang, Z.
Quantum anomalous Hall effect and related topological electronic
states, \emph{Adv. Phys.} \textbf{64}, 227-282 (2015).

\bibitem{N.Marzari1997} Marzari, N. \& Vanderbilt, D. Maximally
localized generalized Wannier functions for composite energy bands,
\emph{Phys. Rev. B} \textbf{56},12847-12865 (1997).

\bibitem{I.Souza2001} Souza, I., Marzari, N. \& Vanderbilt, D.
Maximally localized Wannier functions for entangled energy bands,
\emph{Phys. Rev. B} \textbf{65}, 035109 (2001).

\bibitem{A.A2008} Mostofi, A. A., Yates, J. R., Lee, Y.S.,
Souza, I., Vanderbilt, D.\& Marzari, N. Wannier90: A tool for
obtaining maximally-localised Wannier functions, \emph{Comput. Phys.
Commun.} \textbf{178}, 685-699 (2008).

\bibitem{L.Chaput2011} Chaput, L., Togo, A., Tanaka, I. \& Hug, G.
Phonon-phonon interactions in transition metals, \emph{Phys. Rev. B}
\textbf{84}, 094302 (2011).

\bibitem{Hahn_01959} Hahn, H.\& Ness, P.
\"Uber Subchalkogenidphasen des Titans. \emph{Z. Anorg. Allg. Chem.} \textbf{302}, 17-36 (1959).

\bibitem{Harry1957}Hahn. H., Harder, B., Mutschke, U.\& Ness P.
Zur Kristallstruktur einiger Verbindungen und Phasen des Systems
Zirkon/Schwefel. \emph{Z. Anorg. Allg. Chem.} \textbf{292}, 82-96 (1957).

\bibitem{Steiger1970}Steiger, R.P.\& Cater, E.D. Preparation and identification
of the ZrS phase in the zirconium-sulfur system. \emph{High Temp. Sci.} \textbf{2}, 398-401 (1970).

\bibitem{Hahn1959}Hahn, H. \& Ness, P. \"Uber das System Zirkon/Selen.
\emph{Z. Anorg. Allg. Chem.} \textbf{302 }, 37-49 (1959).

\bibitem{Schewe1994}Schewe-Miller, I. M. \& Young, V. G. Hf$_2$Se$_3$,
a new structure in the binary Hf-Se system. \emph{J. Alloys Compd.} \textbf{216}, 113-115 (1994).

\bibitem{Sodeck1979}Sodeck, H., Mikler, H.\& Komarek, K. L.
Transition metal-chalcogen systems, VI: The zirconium-tellurium
phase diagram. \emph{Monatsh Chem.} \textbf{110}, 1-8 (1979).

\bibitem{G.O2001} \"Orlygsson,G.\& Harbrecht, B.  Structure, properties,
and bonding of ZrTe (MnP type), a low-symmetry, high-temperature
modification of ZrTe (WC type), \emph{J. Am. Chem. Soc.} \textbf{123}, 4168-4173 (2001).

\bibitem{G.O2014} \"Orlygsson,G.\& Harbrecht, B. The crystal structure of
WC type ZrTe. Advantages in chemical bonding as contrasted to NiAs
type ZrTe, \emph{Z. Naturforsch. B} \textbf{54}, 1125-1128 (1999).

\bibitem{add1} Yu, R., Qi, X. L., Bernevig, A., Fang, Z.\& Dai, X.
Equivalent expression of Z$_2$ topological invariant for band
insulators using the non-Abelian Berry connection, \emph{Phys. Rev. B} \textbf{84}, 075119 (2011).

\bibitem{Soluyanov2011}
Soluyanov, A. A., \& Vanderbilt, D., Computing topological
invariants without inversion symmetry, \emph{Phys. Rev. B} \textbf{83}, 235401
(2011).

\end{thebibliography}
\end{document}